\documentstyle[prb,preprint,eqsecnum,aps]{revtex}
\tighten
\begin{document}
\draft
\title{Anderson-Mott transition as a quantum glass problem}

\author{D.Belitz}
\address{Department of Physics and Materials Science Institute,\\
University of Oregon,\\
Eugene, OR 97403}
\author{T.R.Kirkpatrick}
\address{Institute for Physical Science and Technology, and Department of Physics,\\
University of Maryland,\\
College Park, MD 20742}

\date{\today}
\maketitle

\begin{abstract}
We combine a recent mapping of the Anderson-Mott metal-insulator transition
on a random-field problem with scaling concepts for random-field magnets
to argue that disordered electrons near an Anderson-Mott
transition show glass-like behavior. We first discuss attempts to
interpret experimental results in terms of a conventional scaling picture,
and argue that some of the difficulties encountered point towards a glassy
nature of the electrons. We then develop a general scaling theory for a
quantum glass, and discuss critical properties of both thermodynamic and
transport variables in terms of it. Our most important conclusions are
that for a correct interpretation of experiments one must distinguish
between self-averaging and non-self averaging observables, and that
dynamical or temperature scaling is not of power-law
type but rather activated, i.e. given by a generalized Vogel-Fulcher law. 
Recent mutually contradicting experimental results on Si:P are
discussed in the light of this, and new experiments are proposed to test
the predictions of our quantum glass scaling theory.
\end{abstract}
\narrowtext

\section{INTRODUCTION}

\label{sec:I}

The metal-insulator transition that is observed in doped semiconductors and
other disordered solids is not fully understood, despite almost twenty years
of intense experimental and theoretical efforts.\cite{R} There is strong
evidence for both disorder and electron-electron interactions to play an
important role at this transition, which is called an Anderson-Mott
transition (AMT) to distinguish it from a disorder dominated pure
localization or Anderson transition on one hand, and from a correlation
dominated pure Mott transition on the other hand.\cite{Mott} This interplay
between disorder and interactions makes the AMT a very hard problem in
Statistical Mechanics.

On the theoretical side, until very recently virtually all approaches
studied the problem in the vicinity of two dimensions (d=2) by generalizing
Wegner's theory for the Anderson transition.\cite{Wegner79} These theories%
\cite{R} have led to a classification of the AMT into various universality
classes that depend, {\it inter alia}, on the presence or absence of
spin-orbit scattering, magnetic impurities, magnetic fields, etc. For most
of these universality classes, perturbative renormalization group methods
lead to a critical fixed point in $d=2+\epsilon$ dimensions, and standard
critical behavior with power-law scaling is found. However, the framework of
these theories does not allow for an order parameter (OP) description of the
AMT, and does not lead to a simple Landau or mean-field theory. As a result,
the physics driving the AMT remains relatively obscure in this approach,
compared to standard theories for other phase transitions. An alternative
line of attack has recently been explored by the present authors.\cite
{Letter1,Letter2,ZPhys}. In these papers we showed that the same model \cite
{F} that was used for the $2+\epsilon $ expansion allows for an OP
description of the AMT with the tunneling density of states (DOS) as the OP,
and for a simple Landau theory that yields the critical behavior exactly
above the upper critical dimension $d_c^{+}=6$. Furthermore, it was shown
that the problem has random-field aspects and is closely related to a
random-field Ising model. The structure of that theory also suggests that
for certain parameter values, in particular for weak effective
electron-electron interactions, the OP driven AMT can be pre-empted by a
different kind of metal-insulator transition. The most obvious candidate 
is the Anderson transition, where the DOS or OP is uncritical.
These results lead to the suspicion that in large parts of
parameter space important
physical features of the AMT have been missed both in the low-dimensional
theories, and in the interpretations of experiments that were based on these
theories.

On the experimental side, the best studied systems are doped semiconductors,
and some of the most detailed and careful experiments have been done on
Si:P. In what became a benchmark experiment in the field,
Paalanen, Rosenbaum, Thomas and collaborators\cite
{BellLetter1,BellLong,BellRC,BellLetter2} (to be referred to as the Bell
experiment) performed transport measurements at temperatures, $T$, down to $%
2.7\ {\rm mK}$, and used a stress tuning technique to drive a barely
insulating sample through the metal-insulator transition. These experiments
concluded that the critical P concentration, $n_c$, in this system is close
to $3.7\times 10^{18}\,{\rm cm}^{-3}$, and that the conductivity,
extrapolated to $T=0$, vanishes with a critical exponent $s\approx 0.5$,
with error bars of about 3\% for $n_c$ and about 14\% for $s$. The measured
value of $s$ proved hard to understand theoretically, and the critical
behavior of Si:P is still considered enigmatic.\cite{R} More disturbingly,
however, a similar experiment on the same system\cite{K} (to be referred to
as the Karlsruhe experiment) has recently produced results that are
inconsistent with those of Refs.\ \onlinecite{BellLetter1,BellLong,BellRC},
viz. $n_c$ close to $3.5\times 10^{18}\,{\rm cm}^{-3}$, and $s\approx 1.3$.
These disagreements are far greater than the error bars quoted, and have not
been settled between the respective experimental groups.\cite{BellKCommRep}
Measurements of the Hall coefficient have also led to mutually
contradicting results. Dai, Zhang, and Sarachik \cite{SarachikHall} have
reported evidence for a divergent Hall coeficient in Si:P, in contrast to
previous results \cite{KoonCastner} that had found the Hall coefficient to
remain finite. Finally, different experimental results have been obtained
for the crossover exponent of the (longitudinal) conductivity in an external
magnetic field,\cite{magnetoconductivity} but in this case no direct
comparisons on the same system have been done.

The fact that the most careful experiments on the best studied systems lead
to contradictory results is, taken at face value, extremely discouraging. An
important question is what the source of these discrepancies is. One
relevant consideration is the temperature range covered by the respective
experiments, and the lowest temperature reached. This is important since any
measurement of static critical exponents at the AMT involves an
extrapolation to $T=0$. For instance, the discrepancy concerning the Hall
effect results has been blamed in Ref.\ \onlinecite{SarachikHall} on the
earlier experiments not having reached sufficiently low temperatures. In the
case of the conductivity exponent in Si:P, however, extrapolation problems
by themselves are not sufficient to explain the disagreement. References \ %
\onlinecite{BellLetter1,BellLong} and \ \onlinecite{K} agree that the
extrapolated conductivity shows a `tail' at phosphorus concentrations $%
n<3.7\times 10^{18}\,{\rm cm}^{-3}$, but disagree about whether or not this
tail contains the salient physics. The Bell group observed strong
sample-to-sample variations of the conductivity in the tail region at the
lowest temperatures, and concluded that the tail was due to sample
inhomogeneities and should be discarded. The Karlsruhe group, on the other
hand, claims that the tail represents the asymptotic critical region of
the AMT. The fact that no strong sample-to-sample fluctuations were observed
in this case should, however, not be overweighted: The lowest temperature
reached in the Karlsruhe experiment was $T=62.5\ {\rm mK}$, while the strong
fluctuations in the Bell data set in only at lower temperatures. Also,
sample-to-sample variations in the $T$-dependence of the conductivity does
seem to set in just at the lowest $T$ reached in the Karlsruhe experiment, a
phenomenon attributed in Ref.\ \onlinecite{K} to `thermal decoupling', i.e.
problems in reaching and maintaining equilibrium. To summarize the
experimental situation regarding the conductivity in Si:P, one might say
that unusual features, variably described as `rounding', `smearing',
`thermal decoupling', etc. were observed at low temperatures close to the
critical point. These anomalies became stronger upon approaching the
critical point and lowering the temperature, and the main discrepancy
between the Bell and Karlsruhe experiments can be traced to different
assumptions concerning their significance, i.e. whether or not the `tail'
should be taken seriously.

Apart from the conductivity, unusual behavior has also been observed in
thermodynamic properties of doped semiconductors. Both the magnetic
susceptibility, $\chi _m$, and the specific heat, $c_V$, show a pronounced
non-Fermi liqid behavior.\cite{chis,CvBell,CvK} This behavior is observed
near the transition as well as far away from it, is not obviously related to
any critical phenomena near the AMT, and is usually explained in terms of
local moments.\cite{CvBell} However, Bhatt and Fisher \cite{BhattFisher}
have argued that once local-moment/local-moment interactions are taken into
account, one obtains singularities in $\chi _m$ and $c_V$ that are
significantly weaker than those observed experimentally. This suggests that
phenomena other than local moments might contribute to the observed
anomalies in the thermodynamic susceptibilities.

If one insists on a conventional theoretical interpretation of the
conductivity data in terms of power-law scaling\cite{R}, then the
inescapable conclusion is that either the Bell experiment erroneously
discarded the data in the true critical region, or the Karlsruhe experiment
mistook spurious effects for the critical behavior. If this was the case,
then a careful scaling analysis of both data sets and, if necessary, new and
more accurate experiments, should be able to show that the conductivity, and
possibly other quantities, show scaling behavior in one region but not in the
other, thus settling the issue. There is, however, another possibility. If
the theoretical suggestion\cite{Letter2,ZPhys} that the AMT
has random-field
aspects is correct, then one would expect glass-like features and
unconventional scaling similar to what has been predicted\cite
{Villain,Fisher} and observed \cite{Jaccarino} in classical random-field
magnets.

In this paper we explore these possibilities. In Sec.\ \ref{sec:II} we
assume conventional scaling, and check whether a scaling analysis of the
existing data can settle the disagreement between the experimentalists. We
find that it cannot, mostly due to an insufficient temperature range and
the lack of precision experiments that probe the critical behavior of more
than one quantity in a given material. Our analysis suggests, however,
various experiments that might be able to tell which, if any, of the two
doping regions that have been suggested to be the critical one, displays
conventional scaling behavior. In Sec.\ \ref{sec:III} we assume instead that
the AMT features activated scaling of the type found in random-field
magnets, appropriately modified for a quantum problem. Accordingly, we first
develop a general scaling description of a quantum glass, and then work out
predictions for the critical behavior of various observables. We check to
what extent the existing data are consistent with these predictions and
propose new experiments to further investigate this issue.

\section{CONVENTIONAL SCALING}

\label{sec:II}

\subsection{Homogeneity laws}

\label{subsec:II.A}

Let us recall the homogeneity laws for the tunneling or single-particle DOS, 
$N$, and the conductivity, $\sigma$, that express conventional power-law
scaling,\cite{R}

\begin{equation}
N(t,T) = b^{-\beta/\nu} N(tb^{1/\nu},T b^z) \quad,  
\label{eq:2.1}
\end{equation}

\begin{equation}
\sigma (t,T)=b^{-s/\nu }\sigma (tb^{1/\nu },Tb^z)\quad .  
\label{eq:2.2}
\end{equation}
Here $t=n/n_c-1$ denotes the dimensionless distance from the critical point,
and $T$ is the temperature. Since in a quantum problem temperature and
frequency scale the same way,\cite{Tomegafootnote} one
obtains the same homogeneity laws at $T=0$ with $T$ replaced by $\omega $,
where $\omega $ is the external frequency in the case of the conductivity,
and the distance in energy space from the Fermi level, i.e. the bias voltage
in a tunneling experiment, in the case of the DOS. $\beta $ and $s$ are the
critical exponents for the DOS and the conductivity, respectively. $\nu $ is
the correlation length exponent, $z$ is the dynamical critical exponent, and 
$b$ is an arbitrary length scale factor. Analogous homogeneity laws can be
written down for all other quantities of interest.\cite{R} Here we focus on
the DOS, since it is the order parameter for the AMT according to our recent
theory\cite{Letter1,Letter2,ZPhys} and since it is easily measured, and on
the conductivity since it is the most obviously interesting observable in
the context of a metal-insulator transition. Note that throughout this
paper we ignore the possibility of significant corrections to 
scaling.\cite{cts,R}

Putting $b$ equal to the correlation length, $b = \xi \sim t^{-\nu}$, we
obtain

\begin{equation}
N(t,T) = t^{\beta} F_N (T/t^{\nu z}) \quad,  \label{eq:2.3}
\end{equation}
and 
\begin{equation}
\sigma (t,T) = t^s F_{\sigma} (T/t^{\nu z}) \quad,  
\label{eq:2.4}
\end{equation}
where $F_N$ and $F_{\sigma}$ are scaling functions.

These homogeneity laws predict that $N$ and $\sigma$ are functions of a
particular combination of their two arguments $t$ and $T$, and they have
been derived from the traditional description of the AMT.\cite{R} They also
follow from a perturbative treatment of the order parameter theory put
forward recently.\cite{Letter2,ZPhys} As has been mentioned in these
references, it is likely that in the second case the perturbative results
are misleading, and that the dynamical scaling in the physical dimension $%
d=3 $ is of activated rather than of power-law type. We will explore the
consequences of activated scaling in Sec.\ \ref{sec:III} below. For now we
assume that the power-law scaling expressed by Eqs.\ (\ref{eq:2.3}, \ref
{eq:2.4}) is correct asymptotically close to the transition observed in
Si:P. This could be due to either the OP theory of the AMT not being
applicable to Si:P, or to the perturbative analogy between the AMT and
classical random field magnets being misleading. In this section we 
investigate the
experimental consequences of this assumption. In doing so it is important to
realize that the theoretical values of the critical exponents are 
{\it not} known.\cite{exponentfootnote}

Another quantity of interest is the correlation function of the local,
unaveraged DOS, 
\begin{equation}
C(t,T;{\bf x}-{\bf y})=\left\langle N({\bf x})N({\bf y})\right\rangle \quad ,
\label{eq:2.5}
\end{equation}
which according to Refs.\ \onlinecite{Letter1,Letter2,ZPhys} is the order
parameter susceptibility for the AMT. In Eq.\ (\ref{eq:2.5}), $N({\bf x})$
is the local DOS at the Fermi energy, and $<\ldots >$ denotes the impurity
average. The Fourier transform of $C$ obeys the homogeneity law 
\begin{equation}
C(t,T;q)=b^{2+\theta -\eta }\ C(tb^{1/\nu },Tb^z;qb)\quad .  
\label{eq:2.6}
\end{equation}
Here $\eta $ is the usual critical exponent that governs the spatial
dependence of the OP susceptibility, and $\theta \ge 0$ is minus the
scale dimension of a dangerous irrelevant variable in the random-field
problem.\cite{Letter2,ZPhys} In the presence of random-field effects, the
perturbative value of $\theta $ is $2$, while in the absence of random-field
effects one has $\theta =0$. Even though the perturbative result cannot be
correct, at least not in $d=3$, it is very likely that the presence of $%
\theta $ will overcompensate $\eta $, and make the scale dimension of $C$
larger than $2$. This leads to a strong divergence of $C(t=0,T=0;q%
\rightarrow 0)$, the analogue of which has been observed in classical
random-field magnets.\cite{BelangerYoung} Alternatively, one can consider
the homogeneous correlation to find, 
\begin{equation}
C(t,T;q=0)=T^{-(2+\theta -\eta )/z}\ F_C(T/t^{\nu z})\quad ,  \label{eq:2.7}
\end{equation}
with $F_C$ a scaling function. Equation (\ref{eq:2.7}) predicts a strong
divergence as $T\rightarrow 0$, the presence of which can be checked
experimentally. We will come back to this point.

\subsection{Scaling analysis of the Bell experiment}

\label{subsec:II.B}

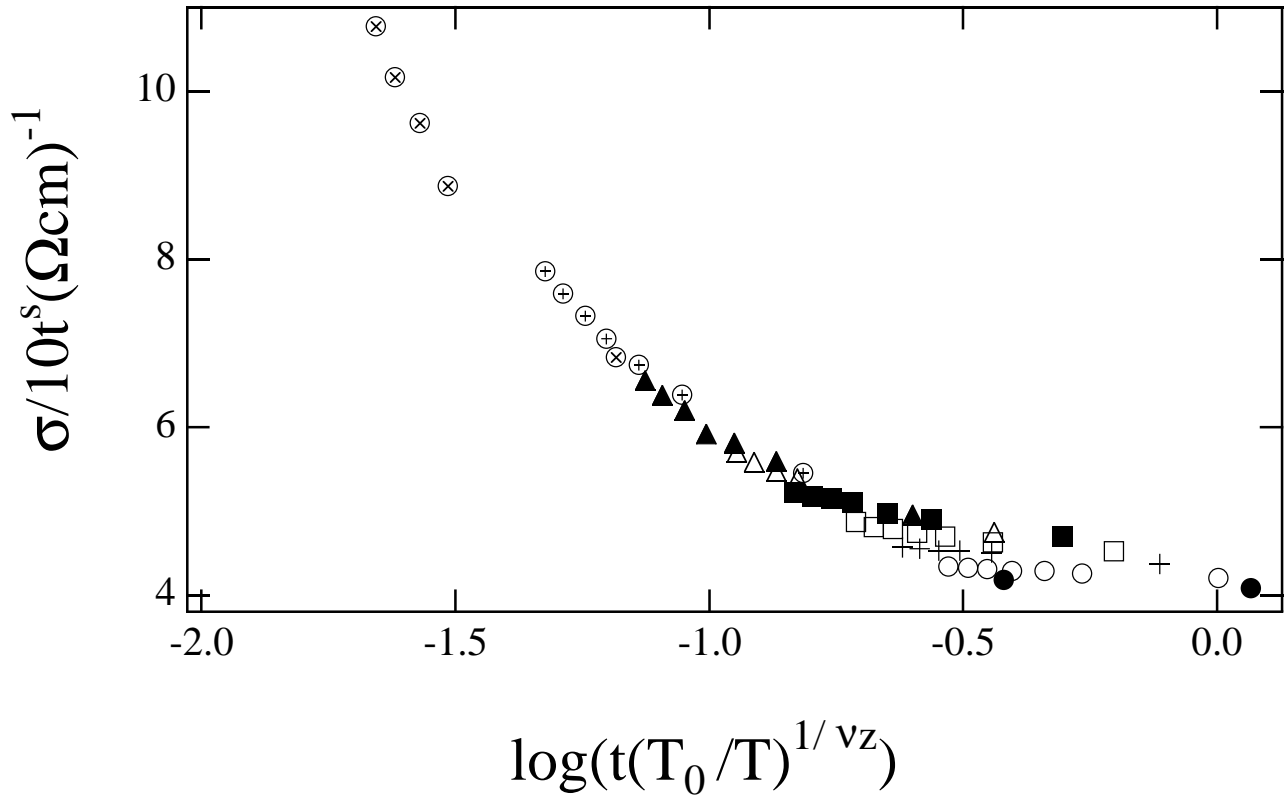
\begin{figure}
\caption{Dynamical scaling plot of the conductivity data from Fig.\ 1 of
 Ref.\ \protect\onlinecite{BellRC}. The plot assumes a critical stress
 $S_c = 6.5\ {\rm kbar}$, and exponent values $s = 0.5$, $\nu z = 2.13$.
 Only data in the temperature range $T < 60\ {\rm mK}$ have been included
 in the plot, and different symbols denote different stress values, from
 $S=6.59\ {\rm kbar}$ to $S=8.03\ {\rm kbar}$. We have chosen 
 $T_0 = 100\ K$, and the relation between $t$
 and $S - S_c$ was taken from Ref.\ \protect\onlinecite{BellLetter1}, viz.
 $t = (S - S_c)\ 5.4 \times 10^{-3}\ ({\rm kbar})^{-1}$.}
\label{fig:1}
\end{figure}

Let us now check whether the data of Ref.\ \onlinecite{BellRC} are
consistent with the above homogeneity law, Eq.\ (\ref{eq:2.4}). From a
double logarithmic plot of the extrapolated $T=0$ conductivity the
experimentalists determined the critical value of the stress, $S_c$, in
their stress-tuning experiment to be $S_c=6.5\pm 0.2\ {\rm kbar}$, and the
conductivity exponent $s\approx 0.5$. From the $T$-dependence of $\sigma $
at a stress value estimated to be close to the critical one, they also
inferred the value of $\nu z\approx 2.7$. In Ref.\ \onlinecite{R} it was
shown that with these exponent values the data do not obey scaling. It must
be emphasized, however, that the error in the value of $\nu z$ is at least
30\%.\cite{BellRC} Accordingly, let us keep $S_c=6.5\ {\rm kbar}$ and $s=0.5$
fixed, but let $\nu z$ float freely to produce the best scaling plot. The
result is shown in Fig.\ \ref{fig:1}, which replots data from Fig.\ 1 of
Ref.\ \onlinecite{BellRC} in a way suggested by Eq.\ (\ref{eq:2.4}). With $%
\nu z=2.13$ one obtains a scaling plot that is better than the one in Ref.\ %
\onlinecite{R}, although its absolute quality is not very good.

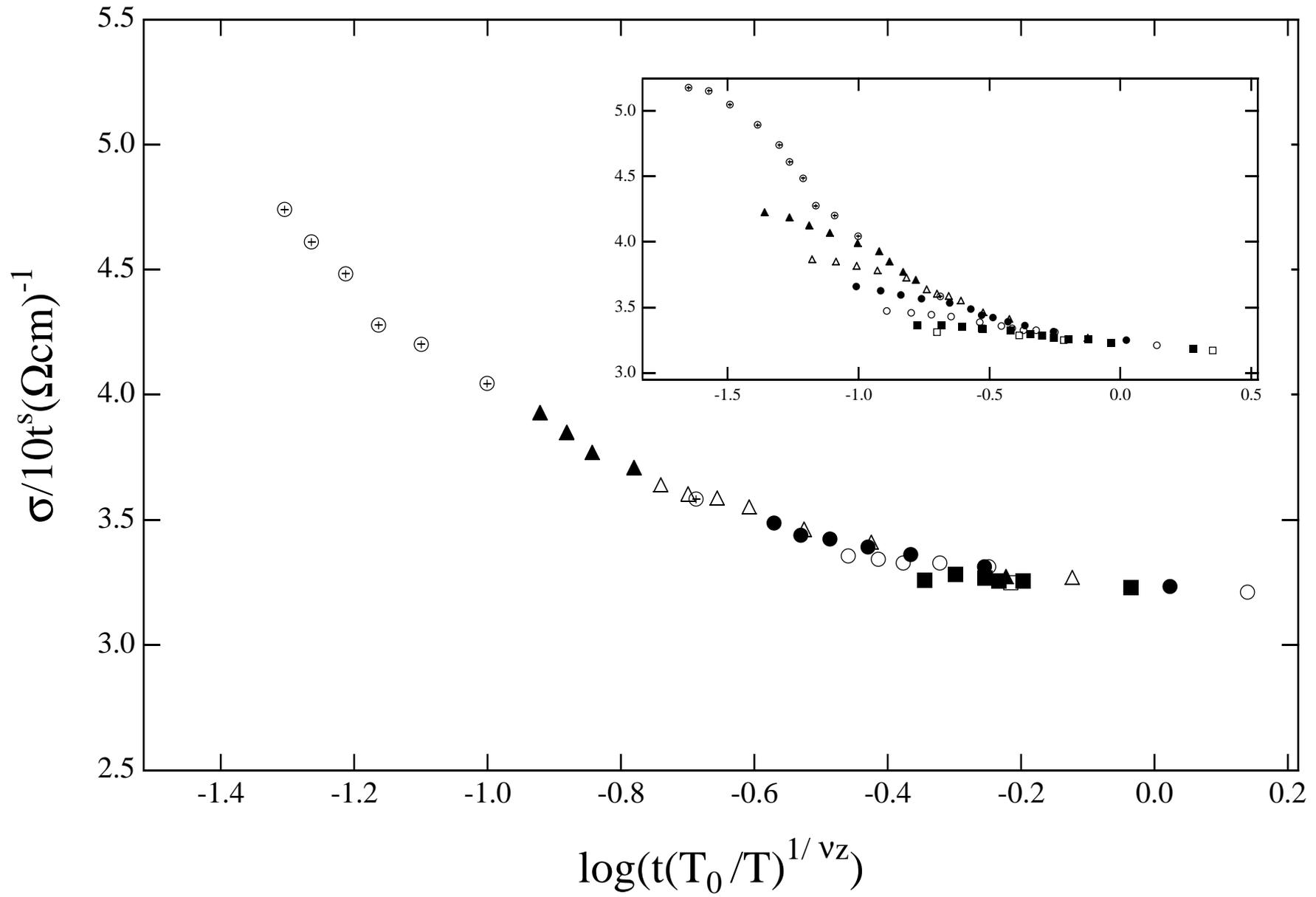
\begin{figure}
\caption{Same as Fig.\ \protect\ref{fig:1}, but with $S_c = 6.71\ {\rm kbar}$,
 $s = 0.29$, and $\nu z = 1.82$. The inset shows that the data cease to scale
 once the temperature region $60\ {\rm mK} < T < 225\ {\rm mK}$ is taken into
 account.}
\label{fig:2}
\end{figure}

We next check whether the quality of the scaling plot can be improved by
changing $S_c$. The largest value of $S_c$ that is consistent with the error
bars given in Ref.\ \onlinecite{BellRC} is $S_c = 6.7\ {\rm kbar}$. Indeed,
inspection of Fig.\ 1 in Ref.\ \onlinecite{BellRC} shows that at $S = 6.59\ 
{\rm kbar}$ the data still show the curvature at the lowest temperatures
that the authors considered characteristic of the insulating regime. Let us
therefore assume that the next higher stress value, $S = 6.71\ {\rm kbar}$,
was the critical one. With this value for $S_c$, we found that the best
scaling plot is achieved with $s=0.29$ and $\nu z = 1.82$, which is shown in
Fig.\ \ref{fig:2}. The quality of the scaling plot is now much better.

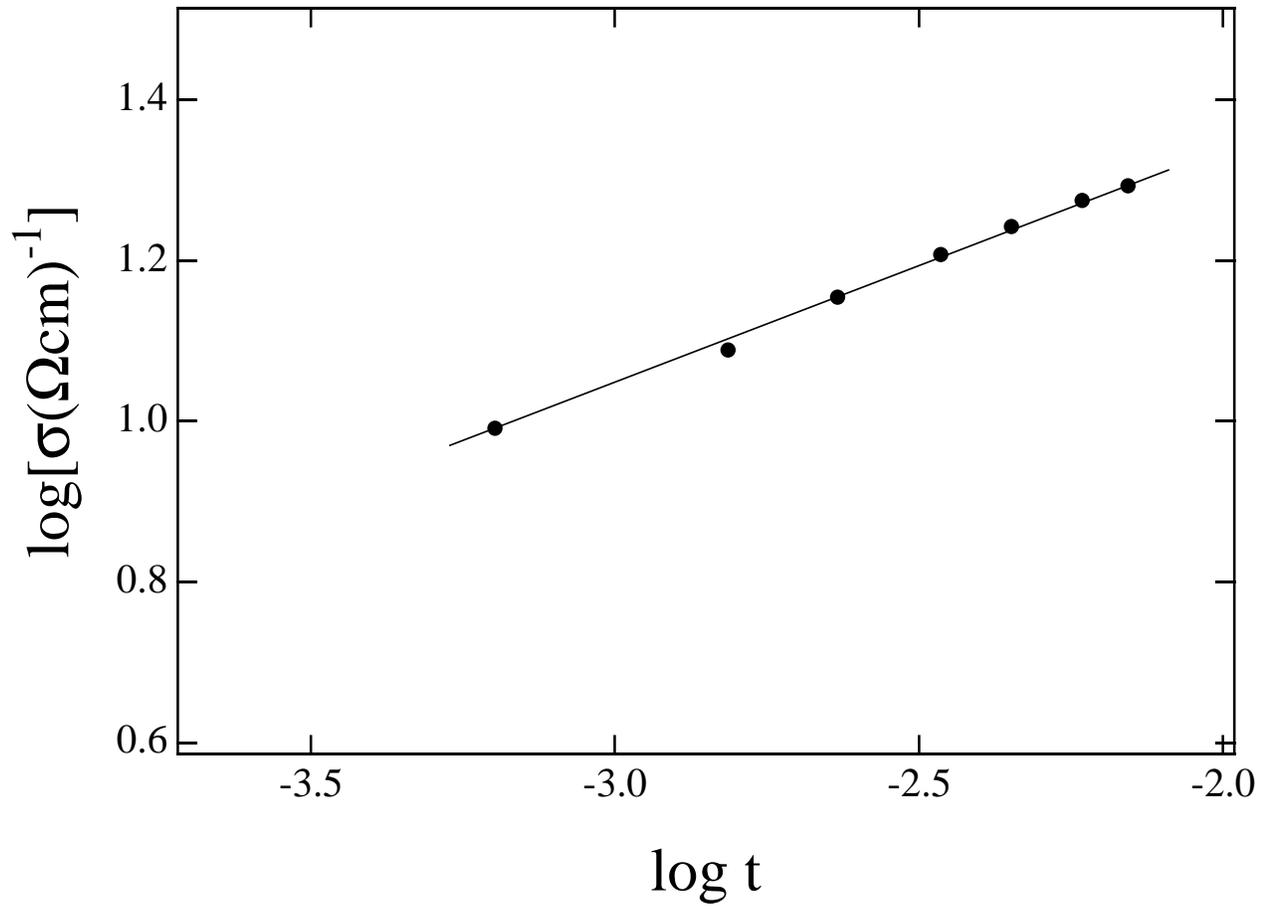
\begin{figure}
\caption{Static scaling plot of the conductivity at $T=0$ with the 
 same parameter values as for the dynamical scaling plot in Fig.\ 
 \protect\ref{fig:2}. The line corresponds to an exponent $s = 0.29$.}
\label{fig:3}
\end{figure}

As Figs.\ \ref{fig:1}, \ref{fig:2} show, the dynamical scaling plot favors a
large value of the critical stress, $S_c$, and a correspondingly small value
of the conductivity exponent $s$. This requires some comments in the light
of the determination $s = 0.51 \pm 0.05$ in Ref.\ \onlinecite{BellRC}. The
determination of $S_c$ and $s$ in Refs.\ %
\onlinecite{BellLetter1,BellLong,BellRC} was guided by the desire to achieve
a good static scaling plot, i.e. a straight line of $\log \sigma$ vs. $\log
t $, over as large a $t$-interval as possible. While this is a legitimate
and often used criterion for determining the critical point, in the case of
Si:P it leads to a rather peculiar result: It is found that an exponent of $%
s=0.5$ fits the behavior of $\sigma (T=0)$ very well out to $t\approx 1.0$.
While this is remarkable and may well have interesting (and presently
unknown) reasons, it is very unlikely that the critical region in Si:P is
that large. Indeed our dynamical scaling plot, which is a more sophisticated
test of scaling than the static one, shows that it is not. On the other
hand, a static scaling plot over a more restricted $t$-range, viz. the data
from Fig.\ 1 of Ref.\ \onlinecite{BellRC} (which is the same data set that
was used to produce the dynamical scaling plots in Figs.\ \ref{fig:1} and 
\ref{fig:2}), shows that $S_c = 6.71\ {\rm kbar}$ and $s=0.29$ fits the data
close to the transition very well, as shown in Fig.\ \ref{fig:3}. This
interpretation suggests a size of the critical region of about 1\%, which is
comparable with the corresponding value for most thermal phase transitions. 
\cite{Sengers}

We conclude that the Bell experiment allows for a reasonably good dynamical
scaling plot, consistent with a conventional power-law scaling
interpretation, provided that the location of the critical point is adjusted
upward, and the value of the conductivity exponent $s$ downward, compared to
the values given in Ref.\ \onlinecite{BellRC}. The resulting small value of $%
s$ aggravates the problem that results from the inequality $\nu \ge 2/3$ in
conjunction with the exponent relation $s=\nu $. This interpretation is
therefore only feasible within theories that allow for $s\neq \nu $, as e.g.
the order parameter description of Ref.\ \onlinecite{Letter2,ZPhys}. Any
theory that yields power-law scaling with $s=\nu $, like
those reviewed in Ref.\ \onlinecite{R}, is inconsistent
with the Bell experiment unless there are large corrections to 
scaling.\cite{cts} For
later reference we also note that the dynamical scaling observed in this
experiment is restricted to a small dynamical range of just a bit over one
decade: For the plot in Fig.\ \ref{fig:2} only data at temperatures $T\leq
60\ {\rm mK}$ were included. Data at higher temperature do {\it not} scale,
as can be seen from the inset in Fig.\ \ref{fig:2}.

\subsection{Scaling analysis of the Karlsruhe experiment}

\label{subsec:II.C}

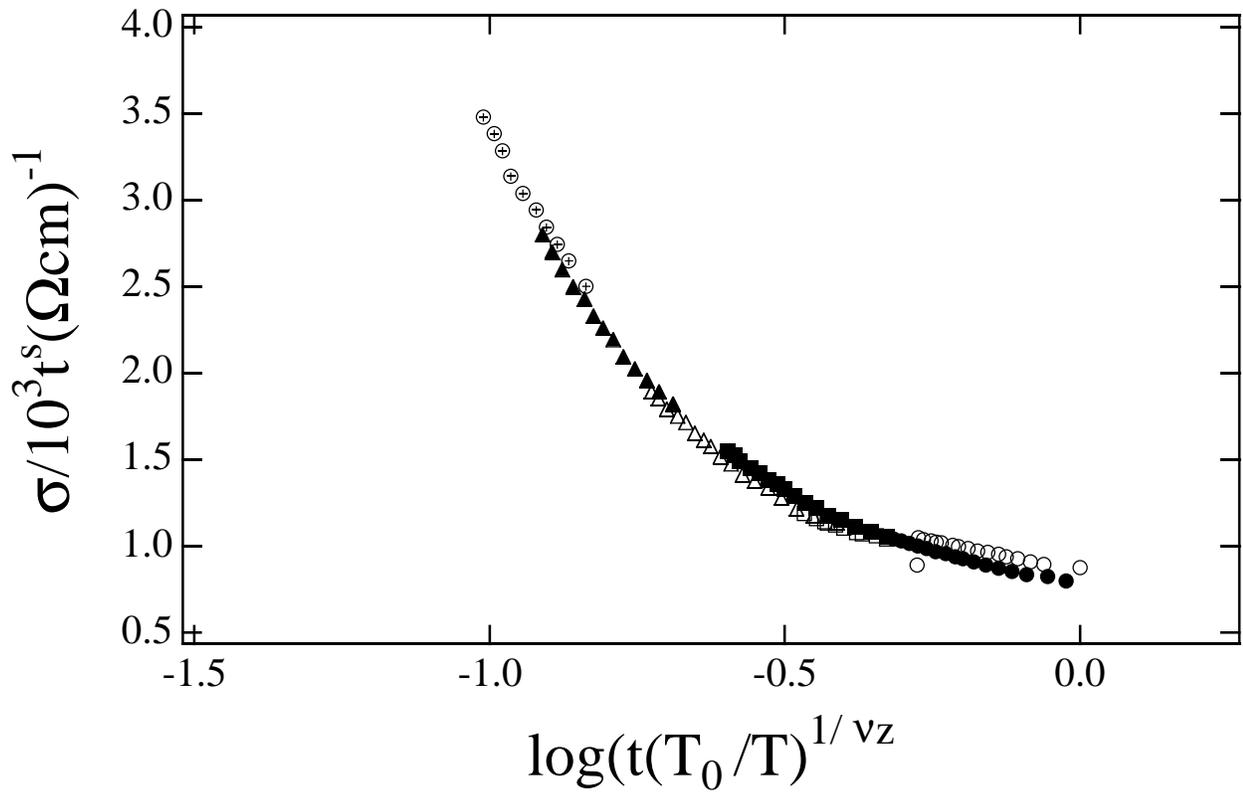
\begin{figure}
\caption{Dynamical scaling plot of the conductivity data from Fig.\ 1 of
 Ref.\ \protect\onlinecite{K}. The plot assumes a critical P density
 $n_c = 3.52\times 10^{18}\ {\rm kbar}$, and exponent values $s = 1.3$,
 $\nu z = 2.7$. Only data in the temperature range $T < 160\ {\rm mK}$ 
 have been included in the plot, and different symbols denote different
 P densities, from $n=3.55\times 10^{18}\,{\rm cm}^{-3}$ to
 $n=3.69\times 10^{18}\,{\rm cm}^{-3}$. We have chosen $T_0 = 100\ {\rm K}$.}
\label{fig:4}
\end{figure}

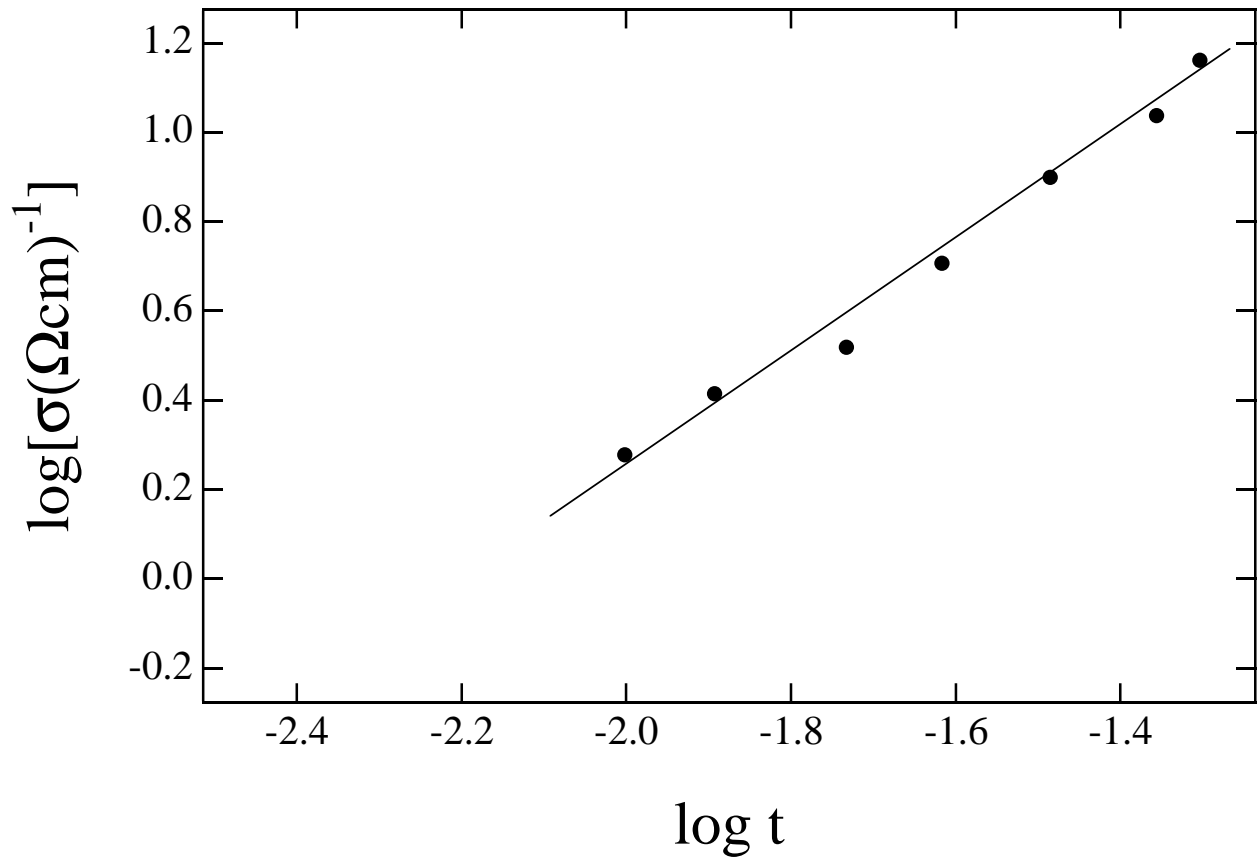
\begin{figure}
\caption{Static scaling plot of the conductivity at $T=0$ with the
 same parameter values as for the dynamical scaling plot in Fig.\
 \protect\ref{fig:4}. The line corresponds to an exponent $s = 1.3$.}
\label{fig:5}
\end{figure}

For the Karlsruhe data a dynamical scaling plot has been given by Stupp et
al. in Ref.\ \onlinecite{BellKCommRep}. For a direct comparison with the
Bell data, we have digitized the data from Fig.\ 1 of Ref.\ \onlinecite{K},
and plot them in Fig.\ \ref{fig:4} in the same way as the Bell data in
Figs.\ \ref{fig:1} and \ref{fig:2}. We have included data for $T < 160\ {\rm %
mK}$, but have left out the lowest temperature points on some samples that
showed obvious rounding effects. We will discuss this rounding in the next
subsection. We have assumed a critical phosphorus concentration of $%
3.52\times 10^{18}\,{\rm cm^{-3}}$, and exponent values $s = 1.3$, and $\nu
z = 2.7$. This yields the plot shown in Fig.\ \ref{fig:4}. The slight
differences between our dynamical scaling plot and that of Stupp et al. (who
found the optimal value of $\nu z$ to be $3.5$) are due to Stupp et al.
optimizing over a larger range of $t$-values, and possibly due to some
errors introduced by redigitizing the data. Overall, however, the two plots
are of comparable quality. Fig.\ \ref{fig:5} shows a static scaling plot
analogous to the one shown in Fig.\ \ref{fig:3}. Again, the assumed value of 
$s$ fits the $T=0$ conductivity well over one decade of $t$.

\subsection{Discussion of the conventional scaling interpretation}

\label{subsec:II.D}

Sections \ref{subsec:II.B},\ \ref{subsec:II.C} and the accompanying figures
show that the data from both the Bell and the Karlsruhe experiment allow for
dynamical scaling plots of equal and satisfactory quality, even though their
results are mutually inconsistent. A check for dynamical scaling is
therefore not sufficient to distinguish between them, and it is necessary to
consider addditional experimental information to settle the issue.

At the heart of the discrepancy lie the different values of $n_c$, which
differ by 6\% between the two groups. An independent measurement of $n_c$
with an error of about 1\% or less would therefore suffice to rule out at
least one of the two interpretations. Unfortunately, both groups claim to
have done just that, but again do not agree on the results. Paalanen et al.%
\cite{BellLetter2} have measured the dielectric polarizability and in at
least one sample have found insulating behavior for values of $n$ as little
as 1\% below their value of $n_c$. Lakner and von L\"ohneysen\cite
{KThermopower} have found the thermopower to exhibit a metallic
characteristic for values of $n$ above their $n_c$, but below that of the
Bell group.

Clearly, what is needed in this situation is an independent determination of 
$n_c$. The existing measurements of the spin susceptibility\cite{chis} and
the specific heat\cite{CvBell} are not suitable for this purpose since they
showed singular behavior in the metallic phase far from the critical point.
One therefore expects these thermodynamic susceptibilities to show a
superposition of critical behavior and some noncritical, but nevertheless
singular, background, which makes them unsuitable for the present purpose.
The most obvious observable that should be free from such complications, and
that is easy to measure, is the DOS. If the conventional scaling scenario is
correct, then the DOS as a function of $t$ and $T$ should show dynamical
scaling, as expressed by Eq.\ (\ref{eq:2.3}), and the quality of the
dynamical scaling plot should be equal to that of the conductivity with the
same parameter values.

Even such an additional measurement, however, may turn out to be
inconclusive unless it is carried out at sufficiently low temperatures, and
unless the nature of the sample-to-sample fluctuations that were observed in
the `tail' region in Refs.\ \onlinecite{BellLetter1,BellLong} is clarified.
The Karlsruhe experiment did not observe such fluctuations, and seemed to
yield a smooth conductivity as a function of doping. However, this may be
misleading since the lowest $T$ reached in this experiment was $62.5\ {\rm mK%
}$, while the sample dependence in the Bell experiment became obvious only
at lower temperature. In this context it is interesting to note that in the
latter there is a clear break in the temperature dependence of the
conductivity around $T=60\ {\rm mK}$, see Fig.\ 1 of Ref.\ %
\onlinecite{BellRC} and the inset in our Fig.\ \ref{fig:2}. Furthermore, in
the same temperature range sample-dependent problems did start to arise in
the Karlsruhe experiment, which were attributed to thermal decoupling, see
Fig.\ 1 of Ref.\ \onlinecite{K}. Finally, we note that all of the work done
by the group at CUNY, both on Si:P\cite{CUNYSiP} and on Si:B \cite{CUNYSiB},
which also yielded a smooth behavior of the conductivity, was at
temperatures higher than $60\ {\rm mK}$.

We conclude that the existing experimental data on Si:P provide evidence
that close to the critical point (whose location is only imprecisely known),
and at temperatures $T\leq 60\ {\rm mK}$ there are large sample-to-sample
fluctuations, and possibly equilibration problems. While the data at higher
temperatures, and the low-temperature data with the `tail' region discarded,
are not inconsistent with conventional scaling behavior, this poses the
question whether the critical behavior at the AMT might be more exotic than
is suggested by Eqs.\ (\ref{eq:2.1}) - (\ref{eq:2.7}). Indeed, our recent
work on an order parameter description of the AMT shows that the AMT has
random-field aspects.\cite{Letter2,ZPhys} Classical random-field magnets are
well known to display glassy behavior with exponentially long equilibration
times, unconventional scaling, etc.\cite{BelangerYoung} It is natural to
assume that similar phenomena can characterize the AMT. While any
conventional scaling interpretation of the AMT in Si:P will necessarily
imply that one of the disagreeing experimental groups made a gross error in
the determination of the critical concentration, an interpretation in terms
of random-field physics has the potential for explaining the unusual
features observed, and the disagreements between the experimentalists, in
terms of real physical effects that are germane to the AMT. We consider this
a very appealing possibility. In the remainder of this paper we therefore
take this suggestion seriously, develop it, and then come back to a
discussion of the experimental situation.

\section{QUANTUM GLASSY BEHAVIOR, AND ACTIVATED SCALING}

\label{sec:III}

\subsection{Scaling theory of a quantum glass transition}

\label{subsec:III.A}

An important characteristic of a glass transition, as opposed to an ordinary
phase transition, is the occurrence of extremely long time scales. While
critical slowing down at an ordinary transition means that the critical time
scale $\tau $ grows like a power of the correlation length, $\tau \sim \xi ^z
$, with $z$ the dynamical critical exponent,\cite{HH} at a glass transition
the critical time scale grows exponentially with $\xi $,\cite{glass} 
\begin{equation}
\ln \ (\tau /\tau _0)\sim \xi ^\psi \quad ,  \label{eq:3.1}
\end{equation}
with $\tau _0$ a microscopic time scale, and $\psi $ a generalized dynamical
exponent. As a result of such extreme slowing down, the system's equilibrium
behavior near the transition becomes inaccessible for all practical
purposes: Even at the smallest feasible frequencies (i.e., the inverse of
the longest feasible waiting times, say, days), finite frequency effects
become noticable (``the systems falls out of equilibrium'') at modest values
of $\xi $ or $t$. It has been proposed that the phase transition in
classical random-field magnets is of this type,\cite{Villain,Fisher} and
experiments have confirmed this conjecture. \cite{Jaccarino} The physical
picture behind this model of random-field magnets is as follows. The
frustration induced by the competition between the exchange interaction and
the random field leads to large clusters of misaligned spins,
that is, locally ordered spins that `point the wrong way', within the
ordered phase. Even though aligning these clusters leads overall to a lower
free energy, it requires a large free energy barrier to be overcome. These
free energy barriers grow like $L^\psi $ as a function of some length scale 
$L$, and near the critical point they diverge like the correlation
length $\xi$ to the power $\psi$. 
The exponent $\psi $ is therefore often referred to as the
`barrier exponent'. Via the Arrhenius law, this leads to Eq.\ (\ref{eq:3.1}).

In a quantum system one expects time and inverse temperature to show the same
scaling behavior, irrespective of whether the critical slowing down follows
an ordinary power law, or 
Eq.\ (\ref{eq:3.1}).\cite{Tomegafootnote} Quantum mechanics thus makes
it even harder to observe the static scaling behavior, since in addition to
exponentially long times or small frequencies it requires exponentially
small temperatures as well. Under realistic experimental conditions the
system will either fall out of equilibrium, or pick up finite temperature
effects. This is a crucial point which we will have to keep in mind for any
discussion of experimental consequences of our ideas.

The role played by temperature in a glassy quantum system can be seen
explicitly in Fisher's recent study of a quantum-mechanical Ising spin chain
in a transverse random magnetic field.\cite{DSFtransverse} This system is
closely related to the classical McCoy-Wu model, for which a number of exact
results have been obtained.\cite{McCoyWu,ShankarMurthy} Since the AMT has
been shown to be a quantum phase transition with random-field aspects\cite
{ZPhys} we believe that these results are qualitatively relevant for our
purposes, and will often use them for comparisons. The physical idea
analogous to the one explained above for spin systems is that while a
repulsive electron-electron interaction always leads to a decrease in
the local DOS, the random potential can in general lead to an increase 
in the local DOS as well.
The competition between these two effects leads to frustration and to, for
example, large insulating clusters within the metallic phase. Delocalizing
these large clusters requires energy barriers to be overcome which again
grow like $\xi ^\psi $ as the transition is approached.

Another question that arises is the scaling behavior of an external magnetic
field. In a spin system the magnetic field, just like the temperature,
obviously sets an energy scale and thus should depend exponentially on the
length scale. In an itinerant electron system, on the other hand, the
magnetic field plays a dual role: It influences the orbital motion of the
electrons, and in that capacity it acts like a length. However, it also
couples to the electron spin via the Zeeman term, and there it acts like an
energy. We therefore expect the slowest 
dependence of a given observable on the
magnetic field to be a logarithmic one, with power-law corrections. A
further consequence of the barrier model is that the frequency or
temperature argument of scaling functions is expected to be
$\ln(\tau/\tau_0)/\ln(T_0/T)$, rather than $\tau T$ as in 
Eqs.\ (\ref{eq:2.1},\ \ref{eq:2.2}). 
The reason is that one expects a very broad distribution
of energy barriers and hence of relaxation times $\tau$. The natural
variable is therefore $\ln\tau$ rather than $\tau$. \cite{logscaling} This
is often referred to as activated scaling.

We next face the question of which observables can be expected to obey
homogeneity laws analogous to the ones given by Eqs.\ (\ref{eq:2.1},\ \ref
{eq:2.2}). First of all, we have to remember that the homogeneity laws hold
for averaged quantities, with the average including both a quantum
mechanical and an impurity or ensemble average, and that we have to
distinguish between self-averaging and non-self averaging quantities.\cite
{Derrida} For the former, their probability distribution for the ensemble
average becomes normal with a vanishing variance in the thermodynamic limit,
so their average value
coincides with the most probable or typical one. For the latter, either the
probability distribution remains broad (e.g. log-normal), or the most
probable value is different from the average one, or both. If the observable
in question is non-self averaging due to a log-normal distribution, then one
expects its logarithm to be a self-averaging quantity.

It is well known that in a system with quenched disorder the free energy is
self-averaging, while the partition function is not, and correlation
functions in general are not, either.\cite{Grinstein,Derrida} Therefore, all
thermodynamic quantities, which can be obtained as partial derivatives of
the free energy, are self-averaging. For a general thermodynamic quantity, 
$Q$, one might therefore expect a homogeneity law 
\begin{equation}
Q(t,T)=b^{-x_Q}\,F_Q\left( tb^{1/\nu }\,,\,{\frac{b^\psi }{\ln (T_0/T)}}%
\right) \quad ,  
\label{eq:3.2}
\end{equation}
where $x_Q$ is the scale dimension of $Q$, 
$F_Q$ is a scaling function, and $T_0$ is a microscopic 
temperature scale, e.g. the Fermi temperature. A
finite frequency will have the same effect as a finite temperature; as
mentioned above, frequency and temperature are interchangeable in a scaling
sense. A magnetic field dependence will be added later, when we discuss the
magnetization.

It is obvious, however, that Eq.\ (\ref{eq:3.2}) can not hold for all
self-averaging quantities, or even for all thermodynamic ones. Suppose $Q$
is some self-averaging quantity, and suppose Eq.\ (\ref{eq:3.2}) holds for 
$Q$. Then $P\equiv Q/T$ is also self-averaging, but since $F_Q$ is not a
homogeneous function of $T$, Eq.\ (\ref{eq:3.2}) will {\it not} hold for $P$.
While this is a rather trivial `breakdown of scaling', it has observable
consequences as we will see below. Here we assume that whether or not a
given thermodynamic quantity obeys Eq.\ (\ref{eq:3.2}) can be decided by
dimensional analysis: If the quantity contains scale dimensions of time or
energy, then it does not obey Eq.\ (\ref{eq:3.2}), else it does. For
instance, the free energy density can not be written in this form, while the
entropy density can, etc. However, the question whether or not a particular
energy acts like an inverse time in a scaling sense can be a nontrivial one.
We will further consider this point when we explicitly discuss various
observables in Sec.\ \ref{subsec:III.B} below.

A rather different class of observables is formed by transport coefficients,
like, e.g., the charge or heat diffusivity. Since they are directly related
to a relaxation time, we expect them to be non-self averaging, while their
logarithms should be self-averaging. Let $\tilde\Xi$ be an {\it unaveraged}
tranport coefficient, i.e. its value for a particular sample or impurity
arrangement. Then we expect its logarithm to obey 
\begin{equation}
\left<\ln\ \tilde\Xi\right>(t,T) = b^{\pm\psi}
F_{\Xi}\left(tb^{1/\nu}\,,\,{\frac{b^{\psi}}{\ln(T_0/T)}}\right)\quad,
\label{eq:3.3}
\end{equation}
with $F_{\Xi}$ a scaling function. Notice that Eq.\ (\ref{eq:3.3})
describes only the leading, i.e. logarithmic, scaling behavior, and neglects
power-law corrections to scaling. The `scale dimension' of 
$\left<\ln\ \tilde\Xi\right>$ is necessarily plus or minus $\psi$, 
with the sign depending
on whether the quantity vanishes or diverges at the transition.

As with conventional scaling at ordinary phase transitions, Eqs.\ (\ref
{eq:3.2},\ \ref{eq:3.3}) hold only for the singular parts of the respective
quantities, and in general there will be nonvanishing, analytic background
contributions. In the case of Eq.\ (\ref{eq:3.3}) another complication is to
be expected. A general transport coefficient $\tilde\Xi$ is related, by
means of an Einstein relation, to the corresponding diffusivity $\tilde\Delta$
via $\tilde\Xi=\tilde\chi\tilde\Delta$, with $\tilde\chi$ an unaveraged
susceptibility. Since $\tilde\chi$ is a thermodynamic quantity,
$\ln\tilde\chi$ is not expected to show scaling behavior. If $\chi$ 
is critical, one therefore expects a
critical, non-scaling background contribution to 
$\left<\ln\tilde\Xi\right>$, in addition to the scaling part 
given by Eq.\ (\ref{eq:3.3}). We will come back to this.

In the above paragraphs we have stated all of the assumptions that enter our
scaling theory of a quantum glass. In the remainder of this paper we
explicitly discuss the behavior of a number of specific observables that are
of interest in the context of the AMT. We work out the consequences of our
assumptions, and compare the results with those obtained from conventional
scaling, Sec.\ \ref{sec:II}, and with the experiments on Si:P discussed
above.

\subsection{Discussion of observables}

\label{subsec:III.B}

In this section we discuss explicitly the consequences of our scaling
assumptions for various observables. We start with thermodynamic quantities,
for which Eq.\ (\ref{eq:3.2}) and the related discussion above are relevant.
Then we turn to the electrical and the thermal conductivity as examples of
transport coefficients which realize the scaling behavior shown in Eq.\ (\ref
{eq:3.3}).

\subsubsection{Tunneling density of states}

\label{subsubsec:III.B.1}

Let us first discuss the tunneling density of states, $N$, as a function of $%
t$ and $T$. We restrict ourselves to the density of states at the Fermi
level, since according to the discussion in connection with Eq.\ (\ref
{eq:3.2}) a finite frequency or energy, which measures the bias voltage or
the distance from the chemical potential, will have the same effect as $T$.
In the theory put forward in Refs.\ \onlinecite{Letter1,Letter2,ZPhys}, $N$
is the order parameter for the AMT, and its scale dimension follows directly
from the order parameter field theory to be $x_N = d-\theta-2+\eta$.
Comparison with Eq.\ (\ref{eq:2.1}) gives the exponent 
relation\cite{Letter2,ZPhys}
\begin{mathletters}
\label{eqs:3.4}
\begin{equation}
\beta ={\frac \nu 2}(d-\theta -2+\eta )\quad .  
\label{eq:3.4a}
\end{equation}
From Eq.\ (\ref{eq:3.2}) we then obtain the generalization of
Eq.\ (\ref{eq:2.1}) to the case of activated scaling as 
\begin{equation}
N(t,T)=b^{-\beta/\nu}F_N\left( tb^{1/\nu }\,,\,{\frac{b^\psi}{\ln (T_0/T)}}
                                                             \right) \quad .  
\label{eq:3.4b}
\end{equation}
\end{mathletters}%
Here $F_N$ is a scaling function, and $\theta $ is an exponent related to a
dangerous irrelevant variable that is characteristic of the random-field
problem.\cite{Grinstein76} In a classical random-field problem thermal
fluctuation are dangerously irrelevant, and $\theta$ is minus the scale
dimension of temperature. In the present context $\theta$ expresses the 
fact that quantum fluctuations are dangerously irrelevant 
at the AMT.\cite{Letter2}

A fundamental question that arises in this context is how many independent
exponents are needed to describe the AMT. In a classical random-field
problem there are three independent 
static exponents, e.g. $\nu$, $\eta$, and $\theta$.\cite{indexfootnote}
In the AMT theory developed in Ref.\ \onlinecite{Letter2,ZPhys} it
turned out that the dynamical exponent $z$ was not independent, reflecting
the irrelevancy of quantum fluctuations, so that there were still three
independent exponents. Here $\psi $ has taken over the role of $z$, and the
question is whether or not it is independent. In order to decide this, let
us recall the physical meaning of the two exponents $\theta$ and $\psi$.
As noted above, $\theta$ is related to a dangerous irrelevant variable
(DIV), $u$, which vanishes as a function of length scale $L$ like
$u\sim L^{-\theta}$. The random-field fixed point is characterized
by $u\Delta$ scaling to a constant, with $\Delta$ the random
potential energy scale.\cite{Grinstein76} 
Hence $\Delta$ must diverge as $L^{\theta}$. 
The free energy landscape of the random-field
problem is a complicated one, with many near-degenerate valleys that are
separated by energy barriers with saddle points. As a function of length
scale $L$, one expects a typical valley elevation to be related to the
random potential and therefore grow like $L^\theta $. The typical saddle point
elevation grows like $L^\psi$, which defines the exponent $\psi$. 
In order for this picture to be
consistent, one must have $\psi \ge \theta $, as pointed out by Fisher in
Ref.\ \onlinecite{logscaling}. Classically, only the valleys contribute to
the free energy, so one expects the exponent characterizing the dangerously
irrelevant thermal fluctuations to be $\theta $, and the barrier exponent to
be $\psi $, and in general the two will be independent. Quantum
mechanically, however, the saddle points also contribute to the free energy,
and one cannot distinguish between the barrier exponent and the exponent
that expresses the fact that quantum fluctuations are dangerously
irrelevant. We therefore expect $\theta =\psi $ in the quantum case,
although in our notation we will continue to distinguish between the two
exponents. This leaves us with three independent exponents, e.g. $\nu $, $%
\eta $, and $\psi =\theta $. A fourth one will be necessary when we discuss
sytems in external magnetic fields in Secs.\ \ref{subsubsec:III.B.4} and 
\ref{subsubsec:III.B.5} below.

As in the case of ordinary power-law scaling, we can eliminate the arbitrary
parameter $b$ from Eq.\ (\ref{eq:3.4b}) and write $N(t,T)$ in two different
scaling forms to emphasize either the static or the dynamic aspects of the
scaling law. The crossover between the two types of
scaling occurs at a temperature $T_{\times }$ for which 
the two arguments of $F_N$ are equal. Ordinarily,
this criterion would lead to a power-law dependence of $T_{\times }$ on $t$,
but activated scaling implies $T_{\times }\sim \exp (-1/t^{\nu \psi })$. As
a result, the static scaling region will be very small unless $\nu \psi $ is
very small.
Since $\nu $ is bounded from below, $\nu \ge 2/d$,\cite
{Chayesetal} this would require $\psi $ to be very 
small.\cite{smallpsifootnote} Whether or not the
static scaling behavior, $N(t,T)\approx N(t,T=0)\sim t^\beta $ is observable
will then strongly depend on the precise value of $\nu \psi $, on the size
of the critical region, and, to a lesser extent, on the value of the
microscopic temperature scale $T_0$. We thus put $b^\psi =\ln (T_0/T)$, and
write Eq.\ (\ref{eq:3.4b}) as 
\begin{equation}
N(t,T)={\frac 1{\left[ \ln (T_0/T)\right] ^{\beta /\nu \psi }}}\ G_N\left[
t^{\nu \psi }\ln (T_0/T)\right] \quad ,  
\label{eq:3.5}
\end{equation}
The scaling function $G_N$ is related to the function $F_N$ in Eq.\ (\ref
{eq:3.4b}) by $G_N(x)=F_N(x^{1/\nu \psi },1)$, and has the properties $%
G_N(x\rightarrow \infty )\sim x^{\beta /\nu \psi }$, and $G_N(x\rightarrow
0)\rightarrow {\rm const}$.

Equation (\ref{eq:3.5}) makes a qualitative prediction that can be used to
check experimentally for glassy aspects of the AMT: Measurements of the
tunneling density of states very close to the transition should show an
anomalously slow temperature depedence, i.e. $N$ should vanish as some power
of $\ln T$ rather than as a power of $T$. While this is a straightforward
check in principle, in practice it may require a very large $T$-range to
distinguish between the two possibilities.
For instance, in classical magnets the frequency had to be varied over seven
decades in order to convincingly demonstrate the presence of activated
scaling.\cite{Jaccarino} However, measurements over a smaller dynamic range
would also be of interest, since they would put experimental bounds on
possible values of $\psi$. This is particularly important since in the
absence of any information about the value of $\psi$ it is impossible to
tell whether at a given temperature one is in the static or the dynamic
scaling regime. Unfortunately, to our knowledge all measurements of $N$
close to metal-insulator transitions have been performed at fixed (and
rather high) temperatures, so that no $T$-dependent data are available for
analysis.

\subsubsection{Order-parameter susceptibility}

We now turn to fluctuations of the order parameter. We first give a
statistical argument that $N$ is indeed, as assumed above, a self-averaging
quantity. This also sheds some light on the crucial role played by the
electron-electron interaction in our theory for the AMT.

Let us write the unaveraged order parameter, $\tilde N$, as its average plus
fluctuations, $\tilde N = N + \delta N$, and consider the mean-square
fluctuation $\left<\left(\delta N\right)^2\right>$. At $T=0$ in a system of
size $L$ one has 
\begin{equation}
\left<\left(\delta N\right)^2\right> = N\,\Phi(t,L)\quad,  \label{eq:3.6}
\end{equation}
with $\Phi$ some function of $t$ and $L$. Far away from the critical point
we expect $\Phi = {\rm const}$, which leads to $\left(\left<\left(\delta
N\right)^2\right>/N^2\right)^{1/2} \sim 1/\sqrt{N}$ as usual. In the
critical region, on the other hand, we expect $\Phi$ to scale, 
\begin{equation}
\Phi(t,L) = b^{\gamma/\nu} \Phi(tb^{1/\nu}, Lb^{-1})\quad,  \label{eq:3.7}
\end{equation}
with $\gamma$ the critical exponent for the order parameter susceptibility.
At criticality, Eq.\ (\ref{eq:3.7}) implies $\left<\left(\delta
N\right)^2\right> \sim N\,L^{2-\eta} \sim L^{d+2-\eta}$, where we have used
the exponent relation $\gamma = \nu (2-\eta)$, and $N \sim L^d$. For the
root mean-square order parameter fluctuations this means 
\begin{equation}
\left(\left<\left(\delta N\right)^2\right>/N^2\right)^{1/2} \sim
L^{(2-d-\eta)/2} \sim L^{-2\beta/\nu}\quad.  \label{eq:3.8}
\end{equation}
For the last relation in Eq.\ (\ref{eq:3.8}) we have assumed hyperscaling to
be valid. In its absence the argument needs a trivial modification, but the
end result is still given by the far right-hand side of Eq.\ (\ref{eq:3.8}).

Equation\ (\ref{eq:3.8}) implies that order parameter fluctuations at the
critical point become small in large systems provided $\eta >2-d$, or $\beta
>0$ (the first condition depends on hyperscaling, while the second one does
not). This has some interesting consequences. In the noninteracting
localization problem, one has $\eta =2-d$, and $\beta =0$.\cite
{AbrahamsLee,R} As can be seen from the above discussion, this means more
than simply that the order parameter is uncritical in the localization
problem: It indicates that the fluctuations of the density of states are
independent of the system size, and as large as the average. Consequently,
one expects that the density of states in a system of noninteracting
disordered electrons has a very broad distribution, and that the Anderson
transition is pathological from a Statistical Mechanics point of view. All
of this is consistent with explicit studies of the Anderson transition.\cite
{AT} Our order parameter description of the AMT, on the other hand, leads to 
$\beta >0$. Therefore the density of states will be self-averaging in accord
with our assumptions in Sec.\ \ref{subsec:III.A} above, and there are no
obvious obstacles for the description of the AMT in terms of the standard
concepts for continous phase transitions.

We now discuss the order parameter correlation function, which we define as 
\begin{equation}
C(t,T;{\bf x} - {\bf y},\omega) = \left<\tilde N({\bf x},\epsilon_F+%
\omega/2) \ \tilde N({\bf y},\epsilon_F-\omega/2)\right>\quad,
\label{eq:3.9}
\end{equation}
or its spatial Fourier transform, $C(t,T;q,\omega)$. Here 
$\tilde N({\bf x},\epsilon)$ denotes the unaveraged, 
local density of states at energy $\epsilon$. From an analogy with 
Eq.\ (\ref{eq:2.6}) one expects $C$ to
scale, and to show an anomalously strong divergence as $q\rightarrow 0$.
However, in random-field systems the correlation function at nonzero
wavenumber is likely to not be normally distributed,\cite{Derrida} while at $%
q=0$ no such complications are expected to occur. Moreover, in our quantum
system a small-$q$ divergence would be cut off by a finite temperature. We
therefore restrict ourselves to a discussion of the homogeneous correlation
function, i.e. the order parameter susceptibility, which obeys 
\begin{equation}
C(t,T;q=0,\omega) = b^{2+\theta-\eta} F_C\left(tb^{1/\nu}\,,\,{\frac{b^{\psi}%
}{\ln(T_0/T)}}\,;\, {\frac{b^{\psi}}{\ln(\omega_0/\omega)}}\right)\quad.
\label{eq:3.10}
\end{equation}
At criticality, the order parameter susceptibility diverges as the
temperature goes to zero, but only as a power of $\ln T$, 
\begin{equation}
C(t=0,T;q=0,\omega=0) \sim
\left[\ln(T_0/T)\right]^{(2+\theta-\eta)/\psi}\quad.  \label{eq:3.11}
\end{equation}

The local density of states is measurable (on a perfect surface) with a
scanning tunneling microscope, or STM. By measuring the density of states
both by means of a tunnel junction and by means of an STM one can therefore
check whether it is indeed a self-averaging quantity: If it is, then in a
large system fluctuations of the local density of states should be small,
and both the local and the junction measurements should give the same
result. Furthermore, by measuring $\tilde N({\bf x})$ across a sample at
pairs of points with a fixed separation it should be possible to measure the
correlation function $C$, and to check the prediction of Eq.\ (\ref{eq:3.11}%
).

We conclude this subsection by noting that $C$,
at least away from the AMT, contains an uncritical,
`mesoscopic' power-law singularity due to hydrodynamic 
diffusion modes.\cite{AS} 
In order to distinguish this singularity from the one given by
Eq.\ (\ref{eq:3.11}) one can use arguments like those used for 
the classical random field problem with a conserved order 
parameter.\cite{HuseWil} The key idea is to consider
a small but finite wavenumber, $q$, subject to a number of constraints.
First we require a self-averaging quantity which according to the
discussion above Eq.\ (\ref{eq:3.10}) requires $q\xi \ll 1$. Second, we
require the critical part of $C$ to effectively be at $q=0$ which leads to
the restriction $\ell q\ll 1/(\ln T_o/T)^{1/\psi }$, with $\ell $ a
microscopic length on the order of the inverse Fermi wavenumber. Third, the
noncritical mesoscopic contribution should be $T$-independent. This will be
the case if $q\ell \gg (T/T_o)^{1/2}$, since this singularity is due to
diffusion. It can be readily verified that these three conditions can be
simultaneously satisfied.

\subsubsection{Specific heat}

\label{subsubsec:III.B.3}

Let us now consider the entropy density, $s(t,T) = \partial f/\partial T$,
where $f$ is the free energy density. As a thermodynamic quantity, $s$ is
expected to be self-averaging, and since its dimension is that of an inverse
volume (in units chosen such that $k_B = 1$) we can write for the singular
part of $s$, 
\begin{equation}
s(t,T) = b^{-d + \theta} F_s\left(tb^{1/\nu}\,,\,{\frac{b^{\psi}}{\ln(T_0/T)}%
} \right)\quad.  \label{eq:3.12}
\end{equation}
The specific heat is obtained from $s$ by means of a logarithmic derivative
with respect to $T$, $c_V = \partial s/\partial\ln T$, which yields for the
singular part of $c_V$, 
\begin{eqnarray}
c_V(t,T) = b^{-(d-\theta+\psi)}\ F_{c_V} \left(tb^{1/\nu}\,,\,{\frac{b^{\psi}%
}{\ln(T_0/T)}}\right)\quad\quad\ \ \,  \nonumber \\
= {\frac{1}{\left[\ln(T_0/T)\right]^{1+(d-\theta)/\psi}}}\
G_{c_V}\left((T/T_0)^{t^{\nu\psi}}\right)\quad.  \label{eq:3.13}
\end{eqnarray}
Here the scaling functions $G_{c_V}$ and $F_{c_V}$ are related by $%
G_{c_V}(x) = F_{c_V}\left((-\ln x)^{1/\nu\psi},1\right)$.

The fact that the specific heat must vanish at $T=0$ puts constraints on the
possible behaviors of the function $G_{c_V}$ at small values of its
argument. One possibility is $G_{c_V}(x\rightarrow 0)\rightarrow {\rm const}$%
. However, a more natural assumption is that $G_{c_V}(x\rightarrow 0)$
vanishes like a power of its argument. This possibility is realized, e.g.,
in the model studied recently by Fisher.\cite{DSFtransverse} In that case,
the specific heat in the critical region goes like 
\begin{equation}
c_V(t,T) = {\frac{T^{\ {\rm const}\times t^{\nu\psi}}}{\left[\ln(T_0/T)%
\right]^{1+(d-\theta)/\psi}}}\quad.  
\label{eq:3.14}
\end{equation}
If we substitute the exponent values that are appropriate for the model of
Ref.\ \onlinecite{DSFtransverse}, viz. $\nu=2$, $\psi=1/2$, $\theta=0$, $d=1$
and ${\rm const}=2$, then we recover from Eq.\ (\ref{eq:3.14}) Fisher's
result for the transverse random-field Ising chain. The most interesting
aspect of this result is the continuously varying exponent in the numerator,
which leads to successive derivatives of $c_V$ becoming divergent as the
critical point is approached. This behavior means that there is a Griffiths
phase, or rather Griffiths region, away from the critical point within which
certain observables become divergent at various values of $t$.\cite
{Griffiths} Regardless of the behavior of $G_{c_V}(x\rightarrow 0)$, we have
a non-Fermi liquid behavior of the system in a finite region around the
critical point: The specific heat coefficient $\gamma(t,T) \equiv c_V(t,T)/T$
diverges as $T\rightarrow 0$, even away from criticality.

For a discussion of the experimental consequences of Eqs.\ (\ref{eq:3.13},\ 
\ref{eq:3.14}) one must keep in mind that the singular contribution to $c_V$
is additive to the noncritical Fermi liquid background that is linear in $T$%
, and that the singular part will dominate only at sufficiently low
temperatures. Measurements of the specific heat in Si:P\cite{CvBell,CvK}
have indeed observed non-Fermi liquid behavior both near the critical point,
and rather far away from it in either phase. This has been interpreted in
terms of local magnetic moments.\cite{CvBell} The relation between the
local moment and quantum glass pictures is currently unclear. It is
interesting to note that both predict singular behavior of thermodynamic
quantities away from the critical point, and it is conceivable that the
glass picture is related to an interacting local moment description.
Also, from the discussion in Sec.\ \ref{sec:II} we suspect that the lowest
temperatures reached in the Si:P experiments ($\approx 30\ {\rm mK}$) were
not low enough to be in the critical region. Since going to substantially
lower temperatures is not realistic, it would be desirable to have similar
measurements performed on a system with a higher Fermi temperature than
Si:P, which has $T_F\approx 100{\rm K}$ near the critical P concentration.

\subsubsection{Magnetization}

\label{subsubsec:III.B.4}

In order to discuss the magnetization, and the magnetic susceptibility in
the next subsection, we need to add an external magnetic field $H$ to our
discussion. As discussed in Sec.\ \ref{subsec:III.A}, the leading effect of
a magnetic field will come from its coupling to the electron spin, and will
scale the same way as the temperature does, viz. $T\sim \exp (b^\psi )$,
$H\sim \exp (b^\psi )$, both up to multiplicative power-law corrections. In
general one
therefore expects $T/H\sim b^{\phi\psi}$, with $\phi$ an exponent that
characterizes differences in the corrections to scaling of $T$ and 
$H$.\cite{HTfootnote} Dimensionally, the
magnetization $m=\partial f/\partial H$ is an inverse volume times a
temperature divided by a magnetic field, and therefore the
scale dimension of $m$ is $d-\theta-\phi\psi$. Therefore we have, 
\begin{equation}
m(t,T,H)=b^{-d+\theta +\phi\psi}\ F_m\left( tb^{1/\nu }\,,\,{\frac{b^\psi }
                     {\ln (T_0/T)}}\,,\,(H/T)b^{\phi\psi}\,,\,
        {\frac{b^\psi }{\ln (T_0/(T+Hb^{\phi\psi}))}}\right) \quad .
\label{eq:3.15}
\end{equation}
Here the last argument of the scaling function $F_m$ expresses the fact
that the Zeeman energy provided by the magnetic field helps the system
to overcome free energy barriers, and its functional form is
motivated by the fact that the effect of a nonzero $H$ will always be cut
off by a nonzero $T$. However, the reverse is not true, which is why one
still needs the second argument containing only $T$. The third argument
contains the physics due to fluctuations within a given free energy
valley, with no attempts to climb over barriers. 

The most interesting
consequence of Eq.\ (\ref{eq:3.15}) is the leading $H$-dependence of 
$m$ at $T=0$ at criticality, which is 
\begin{equation}
m(t=0,T=0,H)\sim {\frac 1{[\ln (T_0/H)]^{-\phi + (d-\theta)/\psi }}}\quad ,
\label{eq:3.16}
\end{equation}
where we have assumed that $F_m(0,0,\infty,1)$ is a finite number.
If we substitute the exponent values that are appropriate for the model
of Ref.\ \onlinecite{DSFtransverse}, namely $\phi = (1+\sqrt{5})/2$ and
$d$, $\theta$, and $\psi$ as given after Eq.\ (\ref{eq:3.14}), then we
recover Fisher's result for the transverse Ising chain.

The physical interpretation of Eq.\ (\ref{eq:3.16}) is as follows. 
Equation\ (\ref{eq:3.15}) says that the magnetic degrees 
of freedom are glassy, and relax slowly just as the singlet 
or DOS degrees of freedom. If the system is cooled in a magnetic 
field then as the magnetic field is turned off, the magnetization
vanishes slowly as a function of both field and time. This behavior is
reminiscent of that found in random field magnets. The difference is that
here no long range magnetic order develops across the transition. However,
we do expect that experiments that examine the difference between field
cooling and zero-field cooling will be very interesting, as they are in
random-field magnetic systems.\cite{BelangerYoung}

\subsubsection{Magnetic susceptibility}

\label{subsubsec:III.B.5}

So far all quantities we have discussed have been both self-averaging (exept
for the order parameter correlation function at nonzero wavenumber) and
scaling, i.e. they obey homogeneity laws of the type given in Eq.\ (\ref
{eq:3.2}). The magnetic susceptibility, $\chi _m=\partial m/\partial H$, is
clearly self-averaging, but does not obey a homogeneity law for the reasons
explained after Eq.\ (\ref{eq:3.2}). Nevertheless, we can obtain the
functional form of $\chi _m$ by differentiating Eq.\ (\ref{eq:3.15}) with
respect to $H$. The leading behavior of the zero-field susceptibility, which 
is produced by the third argument of the scaling function $F_m$, is
\begin{equation}
\chi_m(t,T)=\left[\ln (T_0/T)\right] ^{-(d-\theta)/\psi + 2\phi}\ 
      {\frac 1T}\ G_{\chi _m}\left( (T/T_0)^{t^{\nu \psi }}\right) \quad .
\label{eq:3.17}
\end{equation}
In an ordinary, power-law scaling scenario one would expect the scaling
function $G_{\chi _m}$ to behave such that at $T=0$ the susceptibility is
finite for $t\neq 0$. Here this is not possible since $\chi _m$ is not a
homogeneous function of $T$. Rather, we conclude that the magnetic
susceptibility will diverge as $T\rightarrow 0$ in a region of finite size
around the critical point. This divergence is power-law with logarithmic
corrections, and the exponent of the power-law is a continous function of $t$,
\begin{equation}
\chi_m(t,T) = {\frac{T^{\,-1\,+\,{\rm const}\times 
                                 t^{\nu\psi}}}{\left[\ln(T_0/T)%
                    \right]^{(d-\theta)/\psi - 2\phi}}}\quad.  
\label{eq:3.17'}
\end{equation}
Clearly, this is another manifestation of the Griffiths phenomenon
discussed above in Secs.\ \ref{subsubsec:III.B.3} 
and \ref{subsubsec:III.B.4}, and again our result
is consistent with that of Ref.\ \onlinecite{DSFtransverse} for the
transverse random-field Ising chain.

As mentioned below Eq.\ (\ref{eq:3.14}), there is some similarity between our
results and those obtained from the local moment picture.\cite{CvBell} Both
lead to a divergent magnetic susceptibility and specific heat coefficient in
the metallic phase, in qualitative agreement with experiments on Si:P and
Si:P,B.\cite{chis,CvBell,CvK} The main difference is that we predict a
critical singularity for either quantity, while the local moment picture
yields thermodynamic anomalies that are decoupled from the AMT. Several
points should be kept in mind, however. First, the coupling, or absence of
it, of local moments to the AMT is an unsolved problem.\cite{R} Second,
Bhatt and Fisher\cite{BhattFisher} have pointed out that interactions
between local moments may considerably weaken the effects found in Ref.\ %
\onlinecite{CvBell}.

\subsubsection{Density susceptibility}

\label{subsubsec:III.B.6}

The thermodynamic density susceptibility, $\partial n/\partial\mu$, is not
directly measurable in a three-dimensional system. However, it is of
interest since it enters the Einstein relation between the electrical
conductivity and the mass diffusion coefficient and can therefore influence
the critical behavior of the conductivity.

As a thermodynamic quantity, $\partial n/\partial \mu =\partial ^2f/\partial
\mu ^2$ is self-averaging, but it is not a scaling quantity in the sense of
Eq.\ (\ref{eq:3.2}) since $f$ is not. $\partial n/\partial \mu $ thus
belongs in the same category as the magnetic susceptibility, namely that of
self-averaging, non-scaling observables. In order to determine the critical
behavior of $\partial n/\partial \mu $, we first note that the chemical
potential $\mu $ in this derivative does {\it not} scale like an energy, but
rather like the correlation length to some power. This can be seen from the
explicit formulation of the order parameter field theory for the AMT,\cite
{ZPhys} whose only dependence on the chemical potential is in the $\mu $%
-dependence of $t$. We thus write 
\[
\partial n/\partial \mu =\partial ^2f/\partial \mu ^2=(\partial ^2f/\partial
t^2)(\partial t/\partial \mu )^2+(\partial f/\partial t)(\partial
^2t/\partial \mu ^2)\quad . 
\]
At an ordinary quantum phase transition one would expect
$f\sim b^{-(d+z)}$. At glassy quantum phase transitions, 
$z$ effectively diverges so that the free
energy does not satisfy a simple homogeneneity law. Instead, one expects,
schematically, $b^{-z}\sim T\sim \exp (-b^\psi )\sim \exp (-1/t^{\nu \psi })$,
i.e., the singular part
of the free energy has an essential singularity in $t$ at zero temperature.
This argument is consistent with the results in Secs.\ \ref
{subsubsec:III.B.3} - \ref{subsubsec:III.B.5} where we showed that the
thermodynamics of our model is qualitatively the same as that of the
transverse random-field Ising chain.\cite{DSFtransverse} At $T=0$, the
latter in turn is equivalent to the model considered by Shankar and Murthy,%
\cite{ShankarMurthy} who found the singular part of the free energy to
behave like $f(t,T=0)\sim t^{1/t}$. We therefore expect this behavior to
qualitatively hold in our case as well, which means that $\partial
f/\partial t$ and $\partial ^2f/\partial t^2$ vanish exponentially as $%
t\rightarrow 0$. $\partial t/\partial \mu $ and $\partial ^2t/\partial \mu
^2 $, on the other hand, can diverge at most like a power of $1/t$. This can
be seen as follows. Let $\mu _c$ be the critical value of $\mu $ at a given
value of the disorder. $t$ must vanish as $\mu \rightarrow \mu _c$, and it
can do so either as a power or as an exponential function of of $\mu -\mu _c$%
. In the latter case all of the derivatives of $t$ with respect to $\mu $
also vanish exponentially, while in the former case they may at most diverge
like a power of $1/(\mu -\mu _c)$. Consequently, the singular part of $%
\partial n/\partial \mu $ at zero temperature must vanish exponentially as $%
t\rightarrow 0$. This is in contrast to the conventional scaling scenario,
which yields a power-law dependence of $\partial n/\partial \mu $ on $t$ in
the framework of the order parameter field theory,\cite{ZPhys} and an
uncritical $\partial n/\partial \mu $ in the $2+\epsilon $ expansion.\cite
{F,R}

In addition to this singular part, one expects in general a nonvanishing
analytic background contribution to $\partial n/\partial\mu$.\cite{ZPhys}
The electrical conductivity will then have the same critical behavior as the
charge or mass diffusivity. However, if in a particular system, or for
particular parameter values, that background contribution should vanish,
then the resulting exponential vanishing of $\partial n/\partial\mu$ will
lead to complications in the scaling description of the conductivity, as was
mentioned in the discussion of Eq.\ (\ref{eq:3.3}) above.

\subsubsection{Electrical conductivity}

\label{subsubsec:III.B.7}

We now turn to the behavior of the electrical conductivity. As explained in
Sec.\ \ref{subsec:III.A}, this is not a self-averaging quantity. We
consider instead the unaveraged conducitivity, $\tilde \sigma $, and define
$l_\sigma \equiv <\ln (\sigma _0/\tilde \sigma )>$, with $\sigma_0$ a
suitable conductivity scale, e.g. the solution of the Boltzmann equation.
According to Eq.\ (\ref{eq:3.3}) and Sec.\ \ref{subsubsec:III.B.6} we expect 
$l_\sigma $ to be self-averaging, and to obey 
\begin{eqnarray}
l_\sigma (t,T) &=&b^\psi F_\sigma \left( tb^{1/\nu }\,,\,{\frac{b^\psi }{\ln
(T_0/T)}}\right) \quad \quad \ \ \,  \nonumber \\
&=&\ln (T_0/T)\ G_\sigma \left( t^{\nu \psi }\ln (T_0/T)\right) \quad .
\label{eq:3.18}
\end{eqnarray}
This holds if the density susceptibility has a nonvanishing uncritical
background contribution, as one usually expects to be the case. If $\partial
n/\partial \mu $ vanishes at criticality, then there will be a critical,
non-scaling background contribution to $l_\sigma $, as explained in
connection with Eq.\ (\ref{eq:3.3}).

As in the case of Eq.\ (\ref{eq:3.13}) or (\ref{eq:3.17}), the behavior of
the scaling function $G_{\sigma}$ for large values of its argument is {\it a
priori} unclear. However, we can use physical arguments to determine it. Let
us define $\Sigma \equiv \sigma_0 \exp(-l_{\sigma})$ as a measure of the
conductivity. If $G_{\sigma}(x\rightarrow\infty)$ vanished faster than 
$1/x$, than $\Sigma$ would approach $\sigma_0$ even for arbitrarily small
$t\neq 0$ as $T\rightarrow 0$. Since $\sigma_0$ is a noncritical quantity,
this is unphysical. On the other hand, if $G_{\sigma}(x\rightarrow\infty)$
vanished more slowly than $1/x$, then $\Sigma(T=0)$ would vanish even for
$t\neq 0$. However, for $t\neq 0$ there are no infinite free energy barriers,
and hence density fluctuations are able to relax and $\Sigma$ must be
nonzero. We therefore conclude that $G_{\sigma}(x\rightarrow\infty) \sim 1/x$%
. This yields 
\begin{mathletters}
\label{subsubsec:III.B.2}
\begin{equation}
\Sigma (t,T=0) \sim \exp(-1/t^{\nu\psi})\quad,  \label{eq:3.19a}
\end{equation}
and 
\begin{equation}
\Sigma (t=0,T) \sim T^{G_{\sigma}(0)}\quad.  \label{eq:3.19b}
\end{equation}
Note that at zero temperature $\Sigma$ vanishes exponentially with $t$, and
that at the critical point $\Sigma$ vanishes like a {\it nonuniversal} power
of $T$.

The conclusion that is most important with respect to the interpretation of
experimental results is that the conductivity, $\tilde \sigma $, is not
self-averaging, while $\ln \tilde \sigma$ is self-averaging and
$<\ln \tilde \sigma >$ scales. While in Sec.\ \ref{sec:II} we have shown that
the existing data for the conductivity do allow for scaling plots, we have
also seen that many `strange' features in the experimental results,
especially in the ultra-low temperature results of the Bell experiment, must
be ignored in order to reach that conclusion. Also, it was necessary to let
all exponents and the position of the critical point float. Within
our current activated scaling scenario there is no reason to believe
that $\tilde\sigma$ would scale if one obtained bounds on
the exponents and on $n_c$ by measuring thermodynamic quantities (which do
scale even under the current scenario) at the same low temperatures as the
conductivity. $<\ln \tilde \sigma >$ does scale, 
but would be hard to measure. In
other words, if activated scaling is present but conventional scaling is
used for analyzing experiments, then better experiments will make things
worse rather than better. Furthermore, $\tilde \sigma $ is predicted to not
be a self-averaging quantity, but to have a broad probability
distribution. Measurements of $\tilde \sigma $ at sufficiently low
temperatures should therefore show large sample-to-sample fluctuations.

We propose that the unusual features that were observed in the experiments
on Si:P, particularly in the ultra-low temperature Bell experiment, and
which we have reviewed in Secs.\ \ref{sec:I} and \ref{sec:II}, are
manifestations of the `glassy' behavior that we have derived above. The fact
that the observed anomalies became stronger at lower temperatures is certainly
consistent with this. The fact that other experiments\cite{R} did not
provide any indications for $\tilde \sigma $ not being a well-behaved
quantity is not a valid counterargument, since they all stayed above, or
barely got below, the $60\ {\rm mK}$ where there is a clear break in the $T$%
-dependence of the conductivity, see Ref.\ \onlinecite{BellRC} and Fig.\ \ref
{fig:2}. In order to further check this proposal, one should measure
thermodynamic quantities, preferably the tunneling density of states,
together with the conductivity at as low temperatures, and over as wide a
temperature range, as possible. A system with a higher Fermi temperature
than doped Si would be advantageous, since it would alleviate the need for
ultralow temperatures. ${\rm Ni(S,Se)}_2$ may be promising in this respect.
In a very recent interesting experiment, Jin et al. have found that,
although conductivity data down to $T=30\ {\rm mK}$ do allow for a
conventional dynamical scaling plot, there are hysteresis effects which may
be indicative of a glass-like behavior of the electrons.\cite{TFR}

\subsubsection{Thermal conductivity}

We finally consider the electronic contribution to the thermal conductivity,
which is the product of the specific heat and the heat diffusivity. For the
same reasons as in the case of the electrical conductivity we expect the
thermal conductivity, $\tilde \kappa $, not to be a self-averaging quantity.
We define $l_\kappa \equiv \left\langle \ln (\kappa _0/\kappa )\right\rangle 
$, with $\kappa _0$ the Boltzmann value, and expect\cite{kappafootnote} 
\end{mathletters}
\begin{equation}
l_\kappa (t,T)=b^\psi F_\kappa \left( tb^{1/\nu }\,,\,{\frac{b^\psi }{\ln
(T_0/T)}}\right) \quad .  \label{eq:3.20}
\end{equation}
This equation can be discussed analogously to Eq.\ (\ref{eq:3.18}) for the
conductivity. If we define $K\equiv \exp (-l_\kappa )$ as a measure of the
thermal conductivity, then we obtain an interesting prediction for the
generalized Wiedemann-Franz ratio $K/\Sigma $ at criticality: 
\begin{equation}
K(t=0,T)/\Sigma (t=0,T)\sim T^{G_\kappa -G_\sigma }\quad ,  \label{eq:3.21}
\end{equation}
with $G_\kappa $ and $G_\sigma $ {\it nonuniversal} numbers (see Eq.\ (\ref
{eq:3.19b})). This is in sharp contrast to the conventional scaling
description of the AMT, which predicts that the Wiedemann-Franz law $\kappa
/\sigma =<\tilde \kappa >/<\tilde \sigma >\,\sim T$ holds even at the
transition.\cite{Cetal}

\section{SUMMARY}
\label{sec:IV}

We conclude by briefly summarizing the results of this paper. We have
employed both conventional and activated scaling scenarios to analyze
experiments on the metal-insulator transition in doped semiconductors,
most notably Si:P. Our main goals were to understand the discrepancies
between different experimental findings, and to work out and analyze
the glassy dynamical features of the transition that are suggested by
recent theoretical advances. 

In Section \ref{sec:II} conventional 
scaling ideas were used to interpret existing
experimental data. The most important conclusions were that existing
experiments for the conductivity are inconsistent with each other 
and that, at least in Si:P at very low temperatures, 
there are large sample-to-sample fluctuations, and possibly
equilibration problems, sufficiently close to the critical point. 

In Sec.\ \ref{sec:III} we assumed that the AMT is a quantum glass 
transition, and we developed a general description of such a transition. 
Our chief results are as follows: 
(1) The specific heat and spin susceptibility are singular as
$T\rightarrow 0$ even in the metallic phase, see 
Eqs.\ (\ref{eq:3.14},\ \ref{eq:3.17'}). These results are consistent
with existing experiments, and the theory given here provides an
alternative to the previous explanation in terms of noninteracting
local moments. (2) The DOS is the order
parameter for the quantum glass transition and it is both self-averaging
and a scaling quantity, see Sec.\ \ref{subsubsec:III.B.1}. At criticality,
it is predicted to vanish logarithmically with temperature,
see Eq.\ (\ref{eq:3.5}). The critical behavior of the OP susceptibility
has also been discussed. (3) The 
electrical conductivity, $\tilde\sigma$, is so broadly distributed that 
it is not a self-averaging quantity, but $\ln \tilde \sigma$ is both 
self-averaging and a scaling
quantity, see Sec.\ \ref{subsubsec:III.B.7}. This result was used
to explain the sample-to-sample fluctuations in $\tilde\sigma$
that were observed in Si:P. In Sec.\ \ref{sec:III} we also 
suggested a number of additional experiments to test
the hypothesis that the AMT is a quantum glass transition.

\acknowledgments

We gratefully acknowledge helpful discussions with David Cohen, Stephen
Gregory, Arnulf Latz, Dave Thirumalai, Thomas Vojta, and Martin Wybourne. 
This work was supported by the NSF under grant numbers DMR-92-09879, 
DMR-92-17496, and DMR-95-10185.

\vfill\eject

\begin{references}
\bibitem{R} For a review, see, e.g., D. Belitz and T.~R. Kirkpatrick, 
 Rev. Mod. Phys. {\bf 66}, 261 (1994).
\bibitem{Mott} N.~F. Mott, {\it Metal-Insulator Transitions}, 
 Taylor \& Francis, (London 1990).
\bibitem{Wegner79} F. Wegner, Z. Phys. B {\bf 35}, 207 (1979).
\bibitem{Letter1} T.~R. Kirkpatrick and D. Belitz, Phys. Rev. Lett. {\bf 73},
 862 (1994).
\bibitem{Letter2} T.~R. Kirkpatrick and D. Belitz, Phys. Rev. Lett. {\bf 74},
 1178 (1995).
\bibitem{ZPhys} D. Belitz and T.~R. Kirkpatrick, Z. Phys. {\bf xx}, xxx (1995).
\bibitem{F} A.~M. Finkel'stein, Zh. Eksp. Teor. Fiz. {\bf 84}, 168 (1983)
 [Sov. Phys. JETP {\bf 57}, 97 (1983)].
\bibitem{BellLetter1} M.~A. Paalanen, T.~F. Rosenbaum, G.~A. Thomas, and
 R.~N. Bhatt, Phys. Rev. Lett. {\bf 48}, 1284 (1982).
\bibitem{BellLong} T.~F. Rosenbaum, R.~F. Milligan, M.~A. Paalanen,
 G.~A. Thomas, R.~N. Bhatt, and W. Lin, Phys. Rev. B {\bf 27}, 7509 (1983).
\bibitem{BellRC} G.~A. Thomas, M. Paalanen, and T.~F. Rosenbaum, Phys. Rev.
 B {\bf 27}, 3897 (1983).
\bibitem{BellLetter2} M.~A. Paalanen, T.~F. Rosenbaum, G.~A. Thomas, and
 R.~N. Bhatt, Phys. Rev. Lett. {\bf 51}, 1896 (1983).
\bibitem{K} H. Stupp, M. Hornung, M. Lakner, O. Madel, and H.~v. L\"ohneysen,
 Phys. Rev. Lett. {\bf 71}, 2634 (1993).
\bibitem{BellKCommRep} T.F. Rosenbaum, G.A. Thomas, and M.A. Paalanen, Phys.
 Rev. Lett. {\bf 72}, 2121 (1994); H. Stupp, M. Hornung, M. Lakner, O. Madel,
 and H.v. L\"ohneysen, Phys. Rev. Lett. {\bf 72}, 2122 (1994).
\bibitem{SarachikHall} P. Dai, Y. Zhang, and M.P. Sarachik, Phys. Rev. B
 {\bf 49}, 14039 (1994).
\bibitem{KoonCastner} D.W. Koon and T.G. Castner, Phys. Rev. Lett. {\bf 60},
 1755 (1988).
\bibitem{magnetoconductivity} S. Bogdanovich, P. Dai, M.P. Sarachik,
 V. Dobrosavljevic, and G. Kotliar, unpublished; T.F. Rosenbaum, S.B. Field,
 and R.N. Bhatt, Europhys. Lett. {\bf 10}, 269 (1989); W.N. Shafarman,
 T.G. Castner, J.S. Brooks, K.P. Martin, and M.J. Naughton, Phys. Rev. Lett.
 {\bf 56}, 980 (1986).
\bibitem{chis} M.~A. Paalanen, S. Sachdev, R.~N. Bhatt, and A.~E. Ruckenstein,
 Phys. Rev. Lett. {\bf 57}, 2061 (1986); Y. Ootuka and N. Matsunaga, J. Phys.
 Soc. Japan, {\bf 59}, 1801 (1990).
\bibitem{CvBell} M.~A. Paalanen, J.~E. Graebner, R.~N. Bhatt, and 
 S. Sachdev, Phys. Rev. Lett. {\bf 61}, 597 (1988).
\bibitem{CvK} M. Lakner and H.~v. L\"ohneysen, Phys. Rev. Lett. {\bf 63},
 648 (1989).
\bibitem{BhattFisher} R.~N. Bhatt and D.~S. Fisher, Phys. Rev. Lett. {\bf 68},
 3072 (1992).
\bibitem{Villain} J. Villain, J. Phys. (Paris) {\bf 46}, 1843 (1985).
\bibitem{Fisher} D.~S. Fisher, Phys. Rev. Lett. {\bf 56}, 416 (1986).
\bibitem{Jaccarino} A.~E. Nash, A.~R. King, and V. Jaccarino, Phys. Rev. B
 {\bf 43}, 1272 (1991).
\bibitem{Tomegafootnote} $T$ and $\omega $ can scale in
 different ways if there are multiple time scales, or if there is a
 quantum-classical crossover exponent for the temperature. The latter case is
 not relevant here since the AMT only occurs at zero temperature, and the
 former possibility we do not consider.
\bibitem{cts} T.~R. Kirkpatrick and D. Belitz, Phys. Rev. Lett. {\bf 70},
 974 (1993).
\bibitem{exponentfootnote} In parts of the experimental community there 
 seems to be a 
 misconception that the exponent values {\it are} known from the $2+\epsilon$
 expansion, and that in particular $s=\nu=1$ in $d=3$ for most universality 
 classes. This is not the case; the theoretical statement is\cite{R} 
 $s=\nu\epsilon = 1 + O(\epsilon)$, or $s=\nu=1 + O(1)$ in $d=3$. This leaves
 the exponent values in $d=3$ completely undetermined, 
 especially since it is known
 that the $2+\epsilon$ expansion is very badly behaved. Virtually the
 only constraint on the exponents in $d=3$ is the rigorous inequality
 $\nu \ge 2/3$, see Ref.\ \onlinecite{Chayesetal}. The $2+\epsilon$
 expansion further yields $s=\nu$ in $d=3$,\cite{R} 
 while in the OP description
 \cite{ZPhys} random-field effects relax this condition, and $s$ and $\nu$
 are independent.
\bibitem{Chayesetal} J. Chayes, L. Chayes, D.~S. Fisher, and T. Spencer, Phys.
 Rev. Lett. {\bf 57}, 2999 (1986).
\bibitem{BelangerYoung} For a review, see, D.~P. Belanger and A.~P. Young, 
 J. Mag. Magn. Mat. {\bf 100}, 272 (1991).
\bibitem{Sengers} J.~M.~H. Levelt Sengers and J.~V. Sengers, in {\it
 Perspectives in Statistical Physics}, edited by H.~J. Ravech\'{e}, North
 Holland (Amsterdam 1982), ch.14.
\bibitem{KThermopower} M. Lakner and H.~v. L\"ohneysen, Phys. Rev. Lett.
 {\bf 70}, 3475 (1993).
\bibitem{CUNYSiP} P. Dai, Y. Zhang, S. Bogdanovich, and M.~P. Sarachik,
 Phys. Rev. B {\bf 48}, 4941 (1993).
\bibitem{CUNYSiB} P. Dai, Y. Zhang, and M.~P. Sarachik, Phys. Rev. Lett.
 {\bf 66}, 1914 (1991).
\bibitem{HH} P.~C. Hohenberg and B.~I. Halperin, Rev. Mod. Phys. {\bf 49},
 435 (1977).
\bibitem{glass} See, e.g., P.~W. Anderson in {\it Ill-Condensed Matter},
 edited by R. Balian, R. Maynard, and G. Toulouse, North Holland (Amsterdam
 1979), p.159. Equation\ (\ref{eq:3.1}) is a generalization of the 
 empirical Vogel-Fulcher law $\ln\tau \sim \xi^{1/\nu}$.
\bibitem{DSFtransverse} D.~S. Fisher, Phys. Rev. Lett. {\bf 69}, 534 (1992);
 Phys. Rev. B {\bf 51}, 6411 (1995).
\bibitem{McCoyWu} B. McCoy and T.~T. Wu, Phys. Rev. {\bf 176}, 631 (1968);
 {\bf 188}, 982 (1969).
\bibitem{ShankarMurthy} R. Shankar and G. Murthy, Phys. Rev. B {\bf 36}, 536
 (1987).
\bibitem{logscaling} See Refs.\ \onlinecite{Villain,Fisher}, and D.~S. Fisher,
 J. Appl. Phys. {\bf 61},
 3672 (1987). It is conceivable, however, that in systems which have barriers
 without a broad distribution the scaling variable could be $T\tau$ even
 though $\tau$ grows exponentially with $\xi$, see 
 D.~S. Fisher and D.~A. Huse, Phys. Rev. B {\bf 38}, 386 (1988).
\bibitem{Derrida} See, e.g., B. Derrida, Phys. Rep. {\bf 103}, 29 (1984).
\bibitem{Grinstein} G. Grinstein, in {\it Fundamental Problems in Statistical
 Mechanics VI}, edited by E.~G.~D. Cohen, North Holland (Amsterdam 1985),
 p.147.
\bibitem{Grinstein76} G. Grinstein, Phys. Rev. Lett. {\bf 37}, 944 (1976).
\bibitem{indexfootnote} Some evidence has been presented for $\theta$ not
 being independent, see M. Gofman, J. Adler, A. Aharony, A.~B. Harris,
 and M. Schwartz, Phys. Rev. Lett. {\bf 71}, 1569 (1993), and references
 therein.
\bibitem{smallpsifootnote} There is some evidence that $\theta$, and
 hence $\psi\geq\theta$, is not 
 anomalously small for random field problems, although $\theta\approx 0.2$
 in a $3-d$ Ising spin glass.\cite{BrayMooreMcMillan} For the (partially) 
 exactly soluble transverse 
 random field Ising chain, $\psi =1/2$,\cite{DSFtransverse} while for the 
 classical random field problem in three dimensions, 
 $\theta\approx 1.5$,\cite{OgielskiHuse} and in two dimensions,
 $\theta=1$.\cite{BrayMoore}
 It is perhaps not accidental that $\theta =d/2$ 
 is consistent with all of these results: In a random potential in
 a volume $V=L^d$, eliminating, say, an insulator cluster in a metallic phase 
 allows for a possible energy gain that is $\sim L^{d/2}$. See also
 Ref.\ \onlinecite{Thirumalai}.
\bibitem{AbrahamsLee} E. Abrahams and P.~A. Lee, Phys. Rev. B {\bf 33}, 683
 (1986).
\bibitem{AT} B.~L. Altshuler, V.~E. Kravtsov, and I.~V. Lerner, Zh. Eksp.
 Teor. Fiz. {\bf 91}, 2276 (1986) [Sov. Phys. JETP {\bf 64}, 1352 (1986)];
 A.~D. Mirlin and Y.~V. Fyodorov, Phys. Rev. Lett. {\bf 72}, 526 (1994).
\bibitem{AS} A.~L. Altshuler and B.~I. Shklovskii, Zh. Eksp. Teor. Fiz.
 {\bf 91}, 220 (1986) [Sov. Phys. JETP {\bf 64}, 127 (1986).
\bibitem{HuseWil} D.~A. Huse, Phys. Rev. B {\bf 36}, 5383 (1987);
 P. Wiltzius, S.~B. Dierker, and S.~B. Dennis, Phys. Rev. Lett. {\bf 62},
 804 (1989).
\bibitem{Griffiths} R.~B. Griffiths, Phys. Rev. Lett. {\bf 23}, 17 (1969).
 See Fisher, Ref.\ \onlinecite{DSFtransverse} for a discussion in a context
 very similar to the present one. In particular the occurence of a Griffiths 
 `phase' does {\it not} mean that the transition is smeared, it is still
 sharp with the critical point at $t=0$.
\bibitem{HTfootnote} S. Sachdev, Z. Phys. B {\bf 94}, 469 (1994) has argued
 that $H$ and $T$ scale exactly the same way if the field $H$ couples to
 a conserved quantity. For the AMT in the presence of spin conservation
 this seems to imply $\phi=0$, although more generally one expects
 $\phi\neq 0$ for quantum glass transitions.
\bibitem{TFR} D.~S. Jin, A. Husmann, Y.~V. Zastavker, T.~F. Rosenbaum,
 X. Yao, and J.~M. Honig, `Test of dynamical scaling at the Anderson-Mott
 transition' (unpublished).
\bibitem{kappafootnote} Since the thermal conductivity depends 
 multiplicatively on the specific heat, there may be a weakly singular
 nonscaling contribution to $l_{\kappa}$. However, the leading scaling
 behavior of $\kappa = \exp(-l_{\kappa})$ is given correctly by 
 Eq.\ (\ref{eq:3.20}).
\bibitem{Cetal} C. Castellani, C. DiCastro, G. Kotliar, P.~A. Lee, and
 G. Strinati, Phys. Rev. Lett. {\bf 59}, 477 (1987).
\bibitem{BrayMooreMcMillan} A.~J. Bray and M.~A. Moore, J. Phys. C {\bf 17},
 L463 (1984); W.~L. McMillan, Phys. Rev. B {\bf 30}, 476 (1984).
\bibitem{OgielskiHuse} A.~T. Ogielski and D.~A. Huse, Phys. Rev. Lett. 
 {\bf 56}, 1298 (1986).
\bibitem{BrayMoore} A.~J. Bray and M.~A. Moore, J. Phys. C {\bf 18}, L927
 (1985).
\bibitem{Thirumalai} T.~R. Kirkpatrick and D. Thirumalai, J. de Physique I,
 {\bf xx}, xxx (July 1995); D. Thirumalai, ``From minimal models to real
 proteins: Time scales for protein folding kinetics'' (unpublished), and
 references therein.
\end{references}
\end{document}